# Maternal Anxiety During Pregnancy and Offspring Information Processing Across Development: A Theory-Guided Predictive-Processing Reappraisal of Two Longitudinal Cohorts

Bea R. H. Van den Bergh[1], Martin G. Frasch[2,3]

[1] Research Group Health Psychology and Leuven Brain Institute, KU Leuven (Catholic University of Leuven), Leuven, Belgium (*bea.vandenbergh@kuleuven.be*)

[2] Institute on Human Development and Disability, School of Medicine, University of Washington, Seattle, WA 98195 (*mfrasch@uw.edu*)

[3] Health Stream Analytics, LLC, Seattle, WA, USA

*Corresponding author*: Bea R. H. Van den Bergh (*bea.vandenbergh@kuleuven.be* )

Address: Faculty of Psychology and Educational Sciences, KU Leuven,

Tiensestraat 102, B-3000 Leuven, Belgium

## Highlights

- Prenatal maternal anxiety relates to selective rather than generalized differences.
- Exposed adolescents and adults differ mainly in endogenous cognitive control.
- Exposed infants show reduced attenuation of repeated standard tones.
- Exposed children show larger LPP to neutral rather than emotional pictures.
- Maternal mindfulness during pregnancy shows the opposite standard-tone association.

## Abstract

Prenatal maternal distress has been associated with differences in offspring brain and behavioural development, yet this literature is rarely framed computationally. Predictive-processing (PP) accounts propose that perception, cognition, and action arise from interactions between incoming information and internally generated predictions, partly governed by precision. We reappraised eleven interrelated publications from two non-clinical prospective cohorts, selected using prespecified criteria (Section 2), to assess whether they form a coherent PP-compatible developmental pattern. Eight Leuven publications spanned adolescence to adulthood. Externally cued response inhibition showed no detectable group differences in modestly powered samples, whereas differences emerged when task goals, stimulus relevance, expected outcomes, and response policies had to be generated or maintained endogenously. At age 28, higher prenatal anxiety was associated with lower radial and mean diffusivity in the left superior posterior corona radiata, lower Vocabulary scores, weaker medial prefrontal–left inferior frontal and prefrontal–left supramarginal gyrus connectivity, and greater unpleasantness of induced dyspnoea. Three Tilburg ERP publications examined infancy and early childhood. Higher prenatal anxiety was associated with increased processing of repeated standard tones, fearful relative to happy vocalizations, and neutral rather than pleasant or unpleasant pictures. Maternal mindfulness showed the opposite association for the repeated standard. Across cohorts, findings formed a coherent cross-modal pattern compatible with differences in how relevance and processing resources are allocated to predictable, threat-related, ambiguous, contextual, feedback-related, and bodily information. Because the publications did not estimate precision, prediction errors, or priors, this is a theory-guided, post-hoc reappraisal that generates discriminating hypotheses rather than a formal computational test.

## 1. Introduction

Computational psychiatry increasingly seeks to explain neurodevelopmental risk in terms of the brain's information-processing principles rather than isolated outcomes or circuits (Friston et al., 2014; Hauser et al., 2019), yet this ambition has rarely been tested against the longitudinal, multimodal literature on prenatal maternal distress. Prenatal maternal psychological distress has repeatedly been associated with variation in offspring neurodevelopment, including brain structure and connectivity, stress regulation, attention, temperament, and later emotional and behavioural functioning (Van den Bergh et al., 2020, 2024), and recent reviews additionally report associations with fetal and infant brain structure and function, particularly in limbic, prefrontal, white-matter, and connectivity-related measures, although findings remain heterogeneous across exposures, developmental stages, and methods (Mandl et al., 2024; Wu et al., 2024). Most of this literature, however, has been organized around particular outcomes or neural systems rather than around common principles of information processing.

Predictive-processing (PP) frameworks propose that perception, cognition, and action emerge from interactions between incoming sensory information and internally generated predictions about the world and the body (Clark, 2013; Friston, 2010, 2018). These predictions are informed by prior beliefs, or priors, while their influence relative to incoming information depends on estimated reliability. Precision refers to this context-dependent weighting and regulates how strongly sensory signals or prediction errors influence updating and action. Altered precision weighting has been proposed in several neurodevelopmental and psychiatric conditions (Adams et al., 2013; Fletcher and Frith, 2009; Sterzer et al., 2018). In autism, theoretical accounts have emphasized inflexible or context-insensitive weighting of prediction errors and an altered balance between sensory evidence and prior expectations (Lawson et al., 2014; Van de Cruys et al., 2014, 2017), while computational work suggests that such differences may be task- and context-dependent rather than reflecting a generic prediction deficit (Arthur et al., 2023).

We examine a parsimonious developmental hypothesis: that variation in maternal anxiety during pregnancy is associated with differences in how the developing brain allocates relevance and processing resources to sensory, affective, contextual, feedback-related, and bodily information, and in how cognitive and behavioural regulation becomes internally organized. We use “relevance” and “processing-resource allocation” as functional descriptions of the empirical pattern. Within PP accounts, their context-dependent influence may be mediated by precision weighting, but the original publications did not estimate this mechanism directly.

Applying this framework retrospectively is useful only if it provides more than new terminology for previously reported associations. The potential contribution lies in examining whether findings originally described in terms of attention, habituation, executive functioning, connectivity, cognition, and bodily-signal processing form a selective developmental pattern across ages and modalities. A PP-informed synthesis can distinguish such a pattern from generalized neurodevelopmental dysfunction, identify which forms of information processing appear most consistently involved, and generate more specific hypotheses for prospective and computational testing. This is particularly relevant to developmental-origins research, which has documented diverse offspring correlates of prenatal maternal distress but has rarely examined whether they reflect common principles governing how information is weighted and used to guide perception and action. Interpretational limitations bearing on this synthesis, including the non-independence of publications within each cohort, sample constraints, and the restriction of both cohorts to non-clinical anxiety levels, are catalogued systematically in Section 6 rather than repeated after each finding below.

Findings from other prenatal-distress cohorts are broadly relevant to this question, including differences in fetal and infant functional connectivity, auditory and sensory processing, attention to affective information, white-matter microstructure, and limbic–prefrontal connectivity (Dean et al., 2018; De Asis-Cruz et al., 2020; Graham et al., 2020; Harvison et al., 2009; Hennessey et al., 2024; Hunter et al., 2012; Posner et al., 2016; Qiu et al., 2015; Scheinost et al., 2017). Interoceptive and autonomic pathways may also be relevant because bodily experience depends on the integration of afferent signals with affective and contextual information (Barrett and Simmons, 2015; Paulus and Stein, 2010). Prenatal anxiety-related exposures have additionally been associated with later differences in autonomic regulation, including reduced maternal and infant heart-rate variability (Braeken et al., 2013). However, these studies differ widely in exposure, age, paradigm, and outcome, and few directly test whether the observed associations reflect a common pattern of context-dependent information weighting.

The present article examines this question across eleven interrelated publications from two deeply phenotyped, prospectively followed cohorts: the Leuven Prenatal Project and the Tilburg Prenatal Early Life Stress (PELS) cohort, hereafter referred to as the Leuven cohort and the Tilburg cohort, respectively. Rather than treating the publications as independent replications, we use their developmental and multimodal complementarity to ask whether selective findings from infancy to adulthood can be organized around common questions: which information is treated as relevant, how task goals and expected outcomes are maintained, how feedback and context influence action, and how bodily signals acquire affective meaning. This cross-publication synthesis could not be provided by any individual report and offers a longitudinal complement to newer studies that test particular PP mechanisms more directly but over narrower developmental periods. This reappraisal does not attempt exhaustive coverage of the broader prenatal-distress neuroimaging literature cited above; we focus on the Leuven and Tilburg cohorts because, to our knowledge, few if any prospective cohorts combine comparably dense longitudinal phenotyping with a comparable developmental span, the Leuven cohort was, internationally, among the first to include prenatal assessment and to follow the same cohort, modest in size, into adulthood.

The Leuven cohort began in 1986 and followed firstborn offspring from fetal life into adulthood. The eight publications considered here span ages 14–28: five behavioural, ERP, and fMRI studies of executive and attentional functioning between ages 14 and 20 (Van den Bergh et al., 2005, 2006; Mennes et al., 2006, 2009, 2020; see Table 2), including a direct contrast of exogenous versus endogenous cognitive control (Mennes et al., 2009), and three studies at approximately age 28 examining white-matter microstructure, resting-state functional connectivity, and interoceptive-affective processing (Van den Bergh et al., 2025; Turk et al., 2023; von Leupoldt et al., 2017; see Table 3).

The Tilburg cohort followed a non-clinical community sample from pregnancy into early childhood. The three publications considered here comprised ERP studies of auditory oddball, emotional face-voice, and affective-picture processing between 9 months and 4 years (van den Heuvel, Donkers, et al., 2015; Otte et al., 2015; van den Heuvel et al., 2018b; see Table 4). The Tilburg research programme developed partly from questions raised by the earlier Leuven findings, while extending the investigation to infancy and early childhood and to more fine-grained electrophysiological measures.

In both cohorts, maternal anxiety was assessed prospectively during pregnancy, although the gestational windows, instruments, and analytic approaches differed. In the eight reappraised Leuven publications, offspring were primarily contrasted according to higher versus low-to-medium maternal State anxiety at 12–22 weeks, whereas the three Tilburg publications analysed maternal anxiety symptoms continuously at 8–14 or 15–22 weeks. Maternal anxiety levels at 12–22 weeks were higher in Leuven than in the corresponding Tilburg distributions, but both cohorts comprised non-clinical community samples (see Supplementary Tables S2 and S3). They therefore cannot evaluate stronger claims that severe or sustained adversity produces entrenched threat-related priors. Nor were the original publications designed as formal PP experiments: attention, arousal, sensory adaptation, feedforward processing, and task engagement remain viable explanations for individual findings. The purpose of the reappraisal is consequently not to relabel every association as predictive processing, but to determine whether PP provides a more specific and internally constrained account of the pattern across publications and generates discriminating predictions for future research.

The broader theoretical framework is developed in a companion paper (Frasch and Van den Bergh, 2026), which proposes that early experience across a spectrum from mild to severe calibrates precision weighting, and considers, as separable extensions, the possible emergence of maladaptive priors under sustained adversity and the energetic costs of maintaining high precision. The present empirical companion addresses only the milder, non-clinical, prenatal end of that spectrum: whether maternal anxiety during pregnancy is associated with differences in the context-dependent allocation of relevance and processing resources and in internally organized regulation, without testing the framework's stronger claims about entrenched priors or its energetic extension.

## 2. Aim, Research Question, and Method of Synthesis

### 2.1 Aim and research question

We ask whether findings across eleven interrelated publications from the Leuven and Tilburg cohorts form a selective developmental pattern associated with maternal anxiety during pregnancy. Specifically, we examine whether differences originally described across sensory, affective, cognitive, neural, and interoceptive domains can be organized in terms of the allocation of relevance and processing resources and the development of internally organized regulation. We distinguish directly demonstrated findings from PP-informed interpretations and identify observations that support, constrain, or could falsify this account.

### 2.2 Selection of publications

This article is a theory-guided, retrospective reappraisal of previously published findings, not a systematic review or meta-analysis; the selection procedure is nevertheless reported here so that the corpus is auditable. Candidate publications were drawn from the complete publication inventories of the Leuven Prenatal Project and the Tilburg PELS cohort (n = 25 Leuven; n = 22 Tilburg). Publications were included if they (i) reported an offspring-level neurocognitive, neural, or interoceptive information-processing outcome (behavioural task performance, event-related potentials, task-related or resting-state fMRI, diffusion MRI, or psychophysical magnitude estimation); (ii) used maternal anxiety or (in one publication) maternal mindfulness alongside anxiety prospectively measured during pregnancy as the predictor; and (iii) appeared in a peer-reviewed journal. Excluded publications and their exclusion reasons are listed in full in Supplementary Table S1 and summarized in Figure 1 (Supplementary Figure S1 repeats it with the Table S1 entry numbers). Publications concerning fetal behavioural state, psychophysiological stress regulation, or psychopathology and socio-emotional development are used as developmental context in Sections 3.6 and 4.5 but do not otherwise contribute to the reappraised corpus. Screening was performed by the first author, who directed both cohort programmes, and verified by the second author; disagreements were resolved by discussion. Applying these criteria, no publication from either cohort reporting an offspring-level information-processing outcome with a prenatal predictor was excluded, including publications whose central analyses returned null results. Overlapping participants across publications within a cohort were handled by treating the publications as interrelated follow-ups rather than independent replications (Section 6). Figure 1 summarizes the selection in an adapted flow diagram (cf. Page et al., 2021).

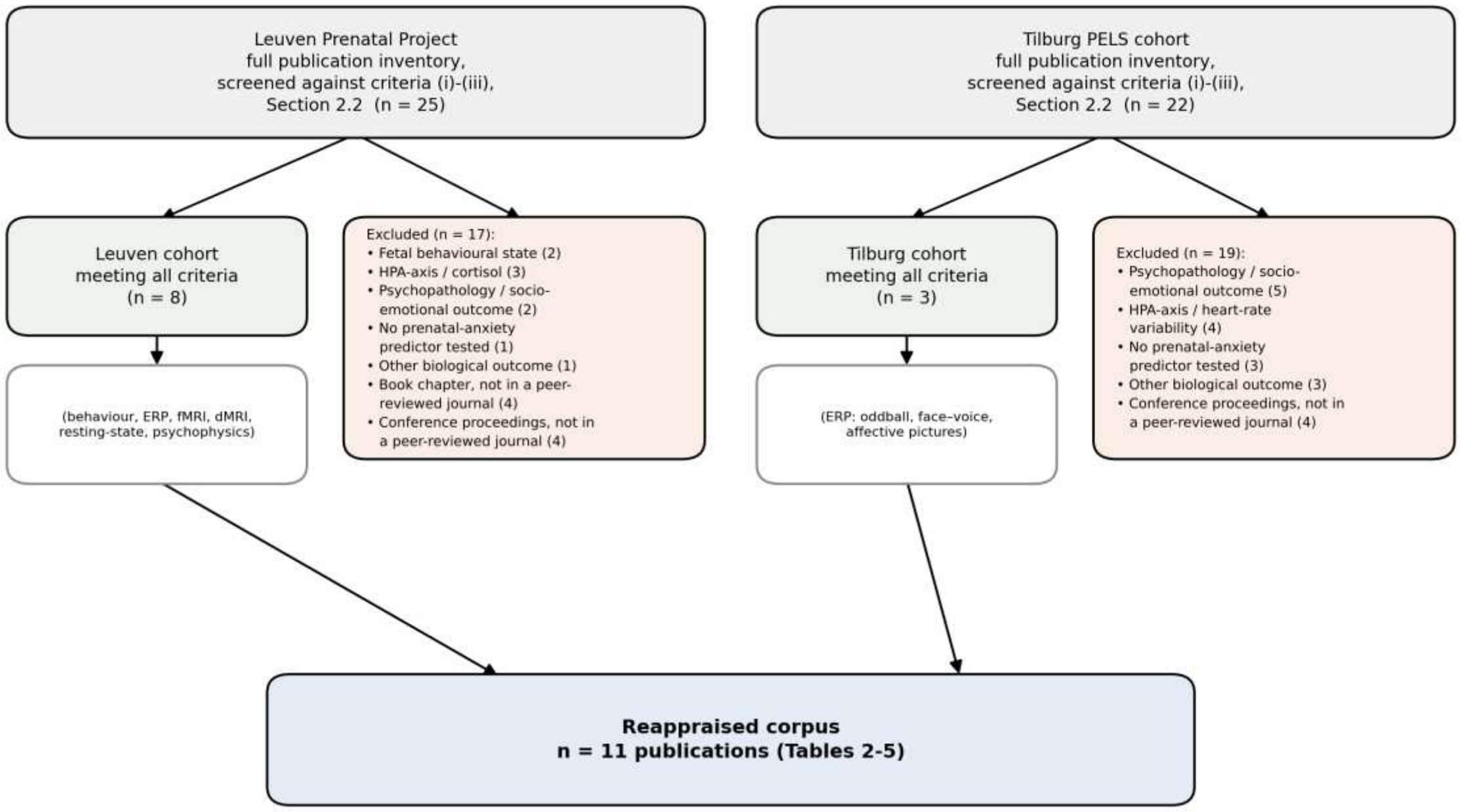

Figure 1. Selection of publications for the reappraisal. Adapted PRISMA-style flow from the full publication inventories of the two cohorts, through the inclusion criteria of Section 2.2, to the eleven reappraised publications (eight Leuven, three Tilburg).

Note. Each publication was assigned a single, primary exclusion reason, i.e., the first of the following checks it failed, in this order: publication venue (not published in a peer-reviewed journal, including book chapters and conference proceedings), whether a maternal-anxiety or maternal-mindfulness predictor was tested, and outcome domain (full exclusion list and reasons in Supplementary Figure S1 and Table S1).

## 2.3 Synthesis approach and evidence tiers

Because the publications are interrelated longitudinal follow-ups rather than independent replications, findings were synthesized narratively and organized by cohort and developmental stage. To make inferential strength explicit, each publication's key findings are audited in Table 5 (Section 6; analysed sample, exposure operationalization, minimum detectable effect at 80% power, and covariate robustness) and Supplementary Table S4 (source statistics, effect-size estimates and uncertainty where retrievable, and evidential-weight tier). Three tiers are used throughout. Tier 1 findings bear most directly on the precision-weighting interpretation: the repeated-standard ERP effect and the endogenous-versus-exogenous control dissociation. Tier 2 findings are compatible with but not specific to it: the emotional face-voice and neutral-picture ERP effects. Tier 3 findings supply possible substrates or distal phenotypes without testing precision weighting: white-matter microstructure, resting-state connectivity, Vocabulary, and dyspnoea unpleasantness. Interpretations are weighted accordingly, and because several central findings rest on null results in modest samples powered to detect only large effects, we describe these outcomes as "no detectable difference" rather than "preserved" throughout: an underpowered null cannot establish that a function is intact, only that no difference large enough to be detected with those samples was found (see Section 6 and Table 5).

## Part I. Leuven Cohort: Endogenous Cognitive Control, Functional Integration, and Interoceptive-Affective Processing

The Leuven cohort examined associations between maternal anxiety during pregnancy and offspring functioning from adolescence into adulthood. The original pregnancy cohort comprised 86 mothers and 88 offspring, including two twin pairs. Maternal State anxiety was assessed at 12–22, 23–31, and 32–40 weeks of gestation. Because scores were not available for all mothers at every assessment, missing trimester-specific values were imputed for the longitudinal analyses. The observed and imputed distributions, and those obtained in the broader Leuven pregnancy dataset, were closely comparable (see Supplementary Table S2).

Higher anxiety was primarily defined as a State Trait Anxiety Inventory (STAI)-State score of 43 or higher at 12–22 weeks, a cutoff corresponding to the 75th percentile in the follow-up samples of the original publications (Van den Bergh et al., 2005; Mennes et al., 2006, 2020) and falling within Decile 7 of the Dutch female reference distribution (Van der Ploeg et al., 1980; see Table 1). One publication (Van den Bergh et al., 2006) used its sample-specific 75th percentile of 43.5, so that its high-anxiety group comprised scores of 44 or higher. Mean scores in the resulting subgroups were 34.4 in the low-to-medium-anxiety group and 49.3 in the high-anxiety group. Depending on the analytic follow-up sample, approximately 25–30% of offspring were classified as prenatally exposed to higher maternal anxiety.

Across most publications, associations with anxiety at 12–22 weeks were examined while accounting for anxiety during the later pregnancy periods and, commonly, postnatal maternal anxiety. Additional maternal, perinatal, and socioeconomic variables were considered where relevant. Together, these publications span executive and attentional processes, task-related and resting-state brain function, white-matter microstructure, cognition, and interoceptive-affective processing, and are reappraised here to test whether these previously separate findings form a more specific pattern involving internally organized regulation and the allocation of relevance and processing resources.

**Table 1. Maternal State anxiety distributions in the Leuven and Tilburg cohorts**

| Cohort | Pregnancy period | Data basis | n | Mean (SD) | Median | P75 | Dutch norm position |
|---|---|---|---|---|---|---|---|

| Cohort | Pregnancy period | Data basis | n | Mean (SD) | Median | P75 | Dutch norm position |
|---|---|---|---|---|---|---|---|
| Leuven | 12–22 weeks | Imputed longitudinal data | 86 | 39.29 (8.54) | 37.50 | 44.00 | Mean D6; P75 D7 |
| Leuven | 23–31 weeks | Imputed longitudinal data | 86 | 34.92 (8.43) | 34.00 | 39.00 | Mean D5; P75 D6 |
| Leuven | 32–40 weeks | Imputed longitudinal data | 86 | 36.45 (7.62) | 36.51 | 40.15 | Mean D5; P75 D6 |
| Tilburg | 8–14 weeks | Observed pregnancy data | 176 | 33.47 (8.97) | 32.00 | 37.25 | Mean D4; P75 D5 |
| Tilburg | 15–22 weeks | Observed pregnancy data | 170 | 32.15 (7.94) | 31.00 | 36.00 | Mean D4; P75 D5 |
| Tilburg | 31–37 weeks | Observed pregnancy data | 153 | 34.23 (8.83) | 34.00 | 38.00 | Mean D5; P75 D6 |

*Note. Leuven values are based on the imputed trimester-specific State anxiety variables used in the longitudinal analyses (n = 86 mothers; 88 offspring, including two twin pairs). Tilburg values are observed scores from the full pregnancy sample. Dutch norm positions are approximate deciles based on the female reference distribution of Van der Ploeg et al. (1980). See Supplementary Table S2 for percentile deciles and clinical threshold counts, and Supplementary Table S3 for cross-cohort statistical comparisons.*

**Table 2. Publications examining endogenous and exogenous cognitive control in the Leuven cohort (Section 3.1)**

| Publication | Age | N | Task / paradigm | Design | Covariates |
|---|---|---|---|---|---|
| Van den Bergh et al. (2005) | 14–15 y | 57 (16 HA, 41 LMA; 29 boys, 28 girls) | Encoding task (present/absent × intact/degraded × low/high memory load); Stop task (go/stop trials, SSRT) | HA vs LMA (between-subjects factor) | Other 2 periods + postnatal anxiety entered; IQ checked post hoc |
| Van den Bergh et al. (2006) | 14–15 y | 64 (16 HA, 48 LMA; 34 boys, 30 girls) | CPT O/Q (time-on-task: block 1 vs. block 6) | HA vs LMA (between-subjects factor); sex-specific effect (males only) | Other 2 periods + postnatal anxiety + parental education + age entered; IQ checked post hoc |
| Mennes et al. (2006) | 17 y | 49 (15 HA, 34 LMA; 29 boys, 20 girls) | Cued Attention, N-back, Go/NoGo, Dual Tasks (Sky Search; Elevator Counting), Response-Shifting (compatible/incompatible × shift/no-shift) | HA vs LMA (between-subjects factor); gender as additional between-subjects factor | Other 2 periods + postnatal anxiety + non-verbal IQ entered |
| Mennes et al. (2009) | 17 y | 23 (8 HA, 15 LMA; boys) | Go/NoGo (exogenous) vs. Gambling (endogenous: gamble/inhibit × determined/underdetermined), ERP | HA vs LMA (between-subjects factor) | Other 2 periods + postnatal anxiety entered |
| Mennes et al. (2020) | 20 y | 17 (8 HA, 9 LMA; men) | Gambling (Go/NoGo/Gamble/Pass trials; exogenous vs. endogenous), task-related fMRI | HA vs LMA (between-subjects factor); extreme-groups subset matched on Performance IQ | Other pregnancy periods, postnatal anxiety, birth weight, parental education checked; none entered (ns) |

*Note. HA = high-anxiety group; LMA = low-to-medium-anxiety group. All studies used maternal State anxiety at 12–22 weeks of gestation as the primary predictor (see Table 1).*

## 3.1 Publications 1–5: Endogenous cognitive control and task-related prefrontal functioning from adolescence to young adulthood

Executive functions are commonly understood as related but partly distinguishable processes supporting goal-directed regulation, including working-memory updating, inhibitory control, and shifting or cognitive flexibility (Miyake et al., 2000). Cognitive control is often used as a closely related term emphasizing the dynamic implementation of goals, task rules, attention, and response selection. In the Leuven studies, endogenous and exogenous control refer not to separate executive-function dimensions, but to different modes through which control is initiated and maintained: internally generated and sustained versus externally triggered by an explicit cue.

Five publications examined executive and attentional processes and their neural correlates between ages 14–15 and 20 years. Across behavioural, electrophysiological, and task-related fMRI assessments, the findings pointed to selective differences in internally generated, or endogenous, cognitive control.

Adolescents whose mothers had reported higher state anxiety during weeks 12–22 of pregnancy responded faster but made more errors in target-absent trials of the Encoding task (see Table 2). This impulsive response pattern reflected a less well-regulated speed-accuracy trade-off; the reaction-time effect remained after controlling for IQ, and the error-rate effect, though attenuated, showed the same pattern in post-hoc comparisons. The pattern was not explained by working-memory load or slower visual scanning, supporting its interpretation as a specific cognitive-regulation difficulty. No corresponding group difference was found in the Stop task, in which an external auditory signal indicated when an ongoing response had to be inhibited. The contrast suggested that externally triggered, or exogenous, response control showed no detectable group differences, whereas greater difficulty emerged when inhibition had to be generated and maintained internally, as in the target-absent Encoding trials, where adolescents had to withhold an initially favoured response long enough to integrate the available information (Van den Bergh et al., 2005).

The association with prenatal maternal anxiety was sex-specific in a subsequent continuous-performance study (see Table 2): boys, but not girls, whose mothers had reported higher anxiety during weeks 12–22 showed a progressive decline in sustained-attention performance as the task continued, becoming slower and more variable over time. This association remained after accounting for maternal anxiety during the other pregnancy periods, postnatal maternal anxiety, parental education, and intelligence, and extended the earlier observations by suggesting difficulty sustaining goal-directed attention and an internally maintained task set over time, although limited to male offspring (Van den Bergh et al., 2006).

Adolescents prenatally exposed to higher maternal anxiety performed more poorly, after adjusting for non-verbal IQ, on tasks requiring the integration and control of several task parameters (see Table 2), whereas working memory, inhibition of a prepotent response, and visual orienting of attention showed no detectable anxiety-related differences; within the study's power limits, this pattern argued against a generalized executive deficit. Based on the cortical map developed for the study, the findings were related to possible subtle developmental differences in the orbitofrontal cortex or in functionally connected cortical and subcortical systems (Mennes et al., 2006).

The distinction between exogenous and endogenous control was next examined using ERPs (see Table 2). Neither behavioural performance nor ERPs during a Go/NoGo task differed according to prenatal maternal anxiety, again showing no detectable difference in externally cued response inhibition. In contrast, differences emerged during a gambling task requiring participants to evaluate probability information, monitor gains and losses, track cumulative score, inhibit competing responses, and determine the most advantageous response without an explicit external instruction. Boys in the high-anxiety group used a less optimal response strategy and showed larger early frontal P2a amplitudes following the gambling stimulus, suggesting the groups differed in how stimulus relevance was established or weighted, for example by focusing more strongly on one salient probability-related feature rather than integrating the complete gambling-stimulus configuration. The effect was confined to prefrontal electrodes, compatible with orbitofrontal or medial prefrontal sources, though reliable source localization was not possible with the available electrode configuration. The convergence of behavioural and electrophysiological differences during gambling, together with their absence during Go/NoGo performance, strengthened the hypothesis of a selective difference in endogenous cognitive control (Mennes et al., 2009).

At age 20, task-related fMRI examined this functional distinction during a gambling paradigm using an extreme-group design matched on Performance IQ (see Table 2). Maternal anxiety during weeks 12–22 was associated with differences across several prefrontal regions rather than an effect confined to the orbitofrontal cortex: in four prefrontal clusters, including the right inferior frontal junction, men in the low-to-medium-anxiety group showed stronger activity during endogenous than exogenous control trials, whereas this task-dependent modulation was absent in the high-anxiety group. No corresponding differences were detected in regions involved in simple response inhibition, consistent with the earlier behavioural and ERP findings, although with eight and nine participants per group this null constrains only very large effects (minimum detectable $d \approx 1.4$ at 80% power; see Table 5). The high-anxiety group also recruited a partly different and more diffuse configuration of prefrontal regions: some regions that deactivated during task performance in the low-to-medium-anxiety group showed little or no comparable deactivation in the high-anxiety group, whereas other regions were recruited only by the high-anxiety group. Together with the less optimal cognitive-control pattern, these findings suggested less efficient endogenous control and possible compensatory recruitment of additional prefrontal resources (Mennes et al., 2020).

Across the five publications, the recurrent finding was not a generalized impairment across executive and attentional domains. Working memory, visual orienting of attention, and externally cued response inhibition showed no detectable group differences, in samples powered to detect only medium-to-large effects (minimum detectable $d \approx 0.8$–1.4; see Table 5). Differences emerged more consistently when cognitive control had to be generated and sustained endogenously, for example, when participants had to maintain task goals, determine the relevance of competing information, monitor previous outcomes, and adapt their response policy without an explicit external cue. The recurrence of this distinction across behavioural tasks, ERPs, and task-related fMRI suggests a coherent difference in how internally maintained goals and task-relevant information are coordinated, providing a direct conceptual bridge to a predictive-processing interpretation.

Viewed through a PP framework, the findings suggest that prenatal maternal anxiety may be associated with lasting differences in how task-relevant information is selected, weighted, and integrated to guide action. Such control requires individuals to infer which information is currently relevant and reliable, maintain expectations about likely outcomes, compare these expectations with feedback, and use the resulting internal model to regulate behaviour. Such internally organized regulation is broadly compatible with active-inference accounts of sequential decision-making, in which agents use internal models to select actions and update beliefs in response to outcomes (Gijsen et al., 2022); however, predictive-processing parameters were not formally manipulated or estimated in the Leuven paradigm. The impulsive speed-accuracy pattern, selective difficulty integrating multiple task parameters, altered processing of task-relevant stimulus features, and absence of typical endogenous-control modulation in prefrontal regions can therefore be interpreted as complementary indications of less efficient internally organized control of task goals, expected outcomes, feedback, and action. This interpretation provides a more specific account than a broad explanation in terms of generalized executive dysfunction.

The timing of the Leuven associations may also be informative. Across several outcomes, maternal state anxiety during weeks 12–22, rather than during later pregnancy, showed the clearest associations. This period overlaps with major neurodevelopmental processes and with differentiation of structures functionally connected with developing prefrontal regulatory systems, including the hippocampus, amygdala, anterior cingulate cortex, and brainstem. The findings should not be localized to an isolated prefrontal region. Instead, they point toward distributed cortical and subcortical systems coordinating sensory evidence, internal goals, expected outcomes, feedback, autonomic regulation, and action. Within such a distributed predictive architecture, medial regions including the ACC may contribute to monitoring conflict, uncertainty, and discrepancies between expected and observed outcomes, whereas lateral and orbitofrontal regions contribute to task-rule implementation, contextual integration, and expected-outcome representations.

Several samples were modest or selective. These constraints, and the broader limits on what these publications can establish about PP mechanisms, are addressed systematically in Section 6. The publications nonetheless provide converging, hypothesis-supporting evidence that can be reformulated within a parsimonious precision-weighting framework.

**Table 3. Publications examining white-matter microstructure, resting-state connectivity, and interoceptive-affective processing at age 28 in the Leuven cohort (Sections 3.2–3.4)**

| Publication | Age | N | Task / paradigm | Design |
|---|---|---|---|---|
| Van den Bergh et al. (2025) | 28 y | 52 (14 HA, 38 LMA; 25 boys, 27 girls); 48 in primary analyses (12 HA, 36 LMA; 24 boys, 24 girls) | Diffusion MRI (DTI, DKI, NODDI), myelin water imaging + WAIS-III (Vocabulary, Block Design, Matrix Reasoning), Trail Making Test-A | HA vs LMA; adjusted for birth weight and postnatal anxiety |
| Turk et al. (2023) | 28 y | 49 (13 HA, 36 LMA; 22 boys, 27 girls) | Resting-state fMRI; ROI-to-ROI connectivity (32 regions) + whole-brain graph measures | HA vs LMA; adjusted for sex, birth weight, postnatal anxiety |
| von Leupoldt et al. (2017) | 28 y | 40 (10 HA, 30 LMA; 20 boys, 20 girls) | Inspiratory threshold-loading; magnitude estimation of dyspnoea intensity/unpleasantness | HA vs LMA |

*Note. HA = high-anxiety group; LMA = low-to-medium-anxiety group. Van den Bergh et al. (2025) and Turk et al. (2023) grouped offspring by maternal State anxiety at 12–22 weeks of gestation (see Table 1); von Leupoldt et al. (2017) grouped offspring by maternal Trait anxiety assessed at the same gestational window.*

## 3.2 Publication 6: White-matter microstructure and cognition at age 28

The white-matter publication examined prenatal maternal anxiety in relation to adult white-matter microstructure and cognitive abilities using diffusion tensor imaging, diffusion kurtosis imaging, neurite orientation dispersion and density imaging, and myelin water imaging, with voxel-wise comparisons between offspring exposed to high versus low-to-medium maternal anxiety during pregnancy (see Table 3). For consistency, most analyses used the 48 participants with complete outcome data; results were unchanged when all available data were included.

Offspring whose mothers reported higher anxiety during pregnancy showed subtle white-matter differences, particularly within a left-lateralized region of the superior posterior corona radiata, with additional trends involving increased fractional anisotropy in the superior anterior corona radiata and higher mean kurtosis in the superior longitudinal fasciculus. Cognitive assessment included the WAIS-III Vocabulary subtest, the Perceptual Organization subtests Block Design and Matrix Reasoning, and the Trail Making Test-A as a measure of processing speed. Perceptual Organization and processing speed showed no detectable group differences (minimum detectable $d \approx 0.9$; see Table 5), whereas Vocabulary scores were lower in the high-anxiety group (Van den Bergh et al., 2025).

Within PP frameworks, long-range white-matter pathways provide the structural infrastructure through which distributed cortical systems exchange sensory evidence, contextual information, and top-down signals (Barrett and Simmons, 2015; Shipp, 2016). The left-lateralized microstructural findings are compatible with subtle developmental variation in distributed information integration rather than generalized cognitive impairment. They may identify structural conditions under which higher-order contextual information is coordinated across regions, complementing the functional differences in internally organized control observed earlier in the cohort.

The accompanying Vocabulary difference is harder to interpret in the same terms. Vocabulary is among the WAIS subtests most sensitive to parental education and the verbal environment, and the adult analyses adjusted for birth weight and postnatal maternal anxiety but not for parental education, which an earlier Leuven publication had included (Van den Bergh et al., 2006). Parental education/socioeconomic status did not differ significantly between the high- and low-to-medium-anxiety groups in this sample (socioeconomic status: $p = .10$; maternal education: $p = .18$; paternal education: $p = .65$; Van den Bergh et al., 2025, Table S1). Because these variables did not differ significantly between groups, their omission makes strong confounding less likely, though it may have reduced statistical power (Miller and Chapman, 2001). Parental education was, nonetheless, broadly associated with cognitive performance in this sample, correlating with Vocabulary and, more modestly, with Perceptual Organization; despite this broader association, the HA/LMA group difference was selective to Vocabulary, with no detectable difference on Perceptual Organization or processing speed. This selectivity is difficult to reconcile with a general socioeconomic pathway, and instead tempers, without eliminating, a socioeconomic account of the Vocabulary finding, particularly given the modest and unequal group sizes. Separately, the original publication reports a modest, statistically uncorrected negative correlation between mean kurtosis in the superior longitudinal fasciculus (a tract implicated in language function) and Vocabulary scores, offering tentative structural support for a microstructural account of the verbal difference; however, this correlation did not survive correction for multiple comparisons and was not observed within the high-anxiety subgroup alone (Van den Bergh et al., 2025).

White-matter indices do not specify the direction or informational content of neural signalling, but they identify a potential substrate through which distributed predictive integration may differ, such as the correlation between SLF microstructure and Vocabulary (although this particular link was statistically weak).

## 3.3 Publication 7: Resting-state functional connectivity at age 28–29

The resting-state publication examined associations between maternal state anxiety during weeks 12–22 and adult functional brain organization using ROI-to-ROI connectivity among 32 cortical and cerebellar regions and whole-brain graph measures, comparing a high-anxiety subgroup with a low-to-medium-anxiety subgroup with adjustment for sex, birth weight, and postnatal maternal anxiety (see Table 3).

Higher prenatal maternal anxiety was associated with weaker functional connectivity involving the medial prefrontal cortex. Network-based statistics additionally identified weaker connectivity involving the left lateral prefrontal cortex, whereas global network measures did not differ significantly between groups (Turk et al., 2023). The selectivity of the findings indicates differences in particular prefrontal connections rather than a generalized reduction in network organization. Parental socioeconomic status did not differ between the high- and low-to-medium-anxiety groups in this cohort (social class based on parental education: $p = .21$; Turk et al., 2023) but was not tested against connectivity outcomes, so an unmeasured association with this specific finding cannot be excluded.

Medial and lateral prefrontal regions participate in distributed systems supporting contextual integration, action selection, and the updating and implementation of higher-order models. Weaker connectivity involving these regions may therefore represent a systems-level correlate of reduced coordination across predictive hierarchies. This interpretation complements the task-related findings: the adolescent and age-20 studies identified atypical prefrontal modulation when endogenous control was required, whereas the resting-state study suggests that differences in prefrontal functional integration remained detectable in adulthood without an explicit task.

The network-based-statistics connection (left lateral prefrontal cortex to left supramarginal gyrus) targets a salience-network node implicated in detecting behaviourally relevant stimuli and coordinating attention with internally generated goals (Turk et al., 2023). This node is the left supramarginal gyrus of the Conn toolbox network atlas, as labelled in Figure 3 of Turk et al. (2023). Salience-network regions more broadly are implicated in relevance detection, attentional reorienting, and dynamic coordination between internally generated and externally driven processing (Andrews-Hanna et al., 2014; Kim, 2014; Menon and D'Esposito, 2022; Costa et al., 2024). This characterization is consistent with the central role assigned to salience-related circuitry in the companion theoretical account of precision-weighted relevance allocation (Frasch and Van den Bergh, 2026).

Resting-state connectivity cannot establish directional message passing or distinguish predictions from prediction errors. Its contribution is to add a network-level correlate, spanning both fronto-associative and salience-network connectivity, to the broader pattern of selective prefrontal and internally organized regulatory differences.

## 3.4 Publication 8: Interoceptive-affective processing at age 28

The interoceptive-affective publication examined dyspnoea perception in healthy adult offspring. Offspring were grouped by maternal Trait rather than State anxiety at 12–22 weeks (see Table 3). Participants completed two magnitude-estimation tasks involving repeated inspiratory threshold loads of increasing resistance. They rated dyspnoea intensity after single inspirations and both intensity and unpleasantness after five successive inspirations. Offspring prenatally exposed to higher maternal anxiety reported greater unpleasantness during the breathing-load challenge. The difference was most evident in the affective evaluation of respiratory sensation rather than as a generalized sensory abnormality and occurred despite normal spirometric lung function and no group difference in concurrent adult state or trait anxiety (von Leupoldt et al., 2017). Parental socioeconomic status was not reported or tested in this publication and cannot be excluded as a contributing factor.

Predictive-processing models of interoception propose that bodily experience arises through the integration of expectations about internal bodily states with incoming afferent signals (Barrett and Simmons, 2015; Paulus and Stein, 2010). Dyspnoea has similarly been conceptualized as emerging from the integration of respiratory sensory signals with expectations and affective context, rather than from respiratory input alone (Harrison et al., 2021; Marlow et al., 2019). From this perspective, the greater unpleasantness reported by the HA group is compatible with altered affective weighting or evaluation of aversive bodily information. The finding broadens the Leuven pattern to interoceptive-affective processing, while not establishing that the same mechanism underlying the cognitive-control findings is involved.

The experiment did not independently manipulate respiratory expectations or sensory reliability and therefore cannot distinguish altered interoceptive precision from differences in appraisal, affective meaning, or prior learning. Its particular contribution is to show that prenatal maternal anxiety was associated with the affective interpretation of bodily challenge almost three decades later, rather than with generalized sensory sensitivity, respiratory pathology, or elevated concurrent anxiety.

## 3.5 Predictive-processing interpretation across the Leuven publications

Across the Leuven publications, higher maternal anxiety during pregnancy was associated with a selective constellation of differences. In the cognitive-control studies, differences were detected for endogenously organized but not for externally cued control (Section 3.1). The adult studies extended the pattern to selected white-matter, connectivity, verbal, and interoceptive-affective outcomes rather than generalized differences across all measures.

Within a PP framework, the most direct evidence concerns internally organized control: behavioural, ERP, and task-fMRI findings converge on differences in selecting and coordinating task-relevant information for action. The adult white-matter and resting-state findings are better treated as possible structural and network-level correlates, and the dyspnoea finding as a distal interoceptive-affective phenotype, rather than as direct tests of precision weighting.

Taken together, the Leuven findings are compatible with developmental variation in how contextual, feedback-related, and bodily information is coordinated, with the strongest support for differences in internally organized cognitive control. They do not establish a specific PP computation, and alternative cognitive and neurodevelopmental explanations remain viable; broader constraints are considered in Section 6.

### 3.6 Relation to developmental psychopathology in the Leuven cohort

Findings from the Leuven longitudinal cohort indicate that higher maternal anxiety during pregnancy was associated with later self-regulatory difficulties and symptoms relevant to developmental psychopathology. Differences were already apparent before birth: maternal anxiety during weeks 12–22 was associated with altered fetal behavioural-state organization (Van den Bergh, 1990), and aspects of fetal sleep organization (conceptually related to this behavioural-state variation, though assessed in a separate publication) were subsequently related to self-regulatory functioning in childhood and adolescence (Van den Bergh and Mulder, 2012). During infancy, exposed offspring showed more regulatory difficulties (Van den Bergh, 1990). At ages 8–9, they reported higher anxiety and showed more parent- and teacher-reported externalizing behavioural problems and ADHD-related symptoms (Van den Bergh and Marcoen, 2004).

In adolescence, prenatal maternal anxiety was additionally associated with HPA-axis dysregulation in male and female offspring. Only in female offspring, however, was this dysregulation associated with reporting more depressive symptoms (Van den Bergh et al., 2008). The selective cognitive and attentional findings from adolescence into young adulthood are discussed in Section 3.1.

Together, the findings are consistent with developmental continuity from altered fetal and infant regulation to later behavioural, stress-regulatory, and emotional difficulties. They do not imply that offspring exposed to higher maternal anxiety generally developed psychopathology or that all outcomes arose through a single pathway. Rather, they suggest increased vulnerability across several domains of self-regulation and developmental psychopathology and provide broader phenotypic context for the more specific cognitive, neural, and interoceptive processes considered above.

## Part II. Tilburg Cohort: Early Sensory and Emotional Processing

The Tilburg Prenatal Early Life Stress cohort followed a non-clinical community sample from pregnancy into early childhood. Maternal psychological characteristics were analysed as continuous predictors and differed across the three publications. They included anxiety symptoms measured with the anxiety subscale of the Symptom Checklist-90, State anxiety measured with the State Trait Anxiety Inventory, and, in the auditory oddball publication, maternal mindfulness measured with the Freiburg Mindfulness Inventory.

For descriptive comparison with Leuven, maternal anxiety in Tilburg fell within a lower decile range of the Dutch female reference distribution in the first half of pregnancy (Leuven 12–22 wk vs. Tilburg 8–14 wk: $d = 0.66$; vs. Tilburg 15–22 wk: $d = 0.88$, both $p < .001$); by late pregnancy the difference, while still statistically significant, was small (Leuven 32–40 wk vs. Tilburg 31–37 wk: $d = 0.26$, $p = .043$) and both cohorts fell within the same decile range (see Table 1; Supplementary Table S3). State anxiety was used as a predictor only in the emotional face-voice publication; the other two publications examined anxiety symptoms measured with the SCL-90.

The three publications comprised event-related potential studies of auditory oddball processing at 9 months, emotional face-voice processing at 9 months, and affective picture processing at age 4 years. Because executive functions are not yet differentiated into adult-like components during infancy and early childhood (Bandettini et al., 2025), the findings are interpreted primarily in terms of sensory, affective, and attentional processing. From a predictive-processing perspective, they are reappraised as possible differences in how predictable, emotionally salient, and affectively uncertain stimuli are assigned relevance and processing resources.

**Table 4. Publications examining auditory, face-voice, and affective-picture processing in the Tilburg cohort (Sections 4.1–4.3)**

| Publication | Age | N | Task / paradigm | Design | Covariates |
|---|---|---|---|---|---|
| van den Heuvel, Donkers, et al. (2015) | 9 mo | 79 infants | Auditory oddball ERP: standard tone + 3 deviant types | Continuous anxiety (SCL-90) and mindfulness predictors | GA, birth weight, postnatal anxiety checked; none entered (ns) |
| Otte et al. (2015) | 9 mo | 82 infants | Face-voice emotional priming ERP: happy/fearful face + vocalization | Continuous anxiety predictor (SCL-90 and STAI-State) | Alcohol, sex, GA, birth weight, age checked: no effect; postnatal anxiety checked: P350 interaction attenuated to trend (p = .043 to .052) |
| van den Heuvel et al. (2018b) | 4 y | 86 children | Affective-picture viewing (IAPS); late positive potential (LPP) | Continuous anxiety predictor (SCL-90) | GA + postnatal anxiety entered; FDR-corrected |

*Note. SCL-90 = Symptom Checklist-90 anxiety subscale; STAI = State Trait Anxiety Inventory. See Table 1 for Tilburg maternal anxiety distributions; GA = Gestational age at birth*

## 4.1 Publication 9: Auditory oddball processing at 9 months

In the auditory oddball publication, maternal anxiety symptoms and mindfulness were assessed at approximately 20.7 weeks of gestation. Anxiety was measured with the anxiety subscale of the Symptom Checklist-90 and mindfulness with the Freiburg Mindfulness Inventory. The two maternal characteristics were examined in separate repeated-measures ANCOVAs, with relevant maternal and perinatal characteristics, including postnatal maternal anxiety, considered as potential covariates.

Infants were presented with a frequent complex standard tone and three types of infrequent deviants: the same tone presented after a shorter interstimulus interval, a white-noise segment, and unique environmental sounds (see Table 4). These deviants respectively introduced temporal irregularity, a marked but non-semantic acoustic change, and varying environmental novelty.

Higher prenatal maternal anxiety was associated with larger N250 amplitudes to the repeated standard tone, whereas no significant associations emerged for any of the three deviant types. Maternal mindfulness during pregnancy showed the opposite association with the standard response: higher mindfulness was related to smaller N250 amplitudes and larger P150 amplitudes (van den Heuvel, Donkers, et al., 2015). Infant electrophysiological responses to expected and unexpected events have been proposed as potential indicators of predictive processing and prediction-error signalling (Berger and Posner, 2023). In the present publication, however, the anxiety-related association concerned the repeated standard rather than the deviants. Within a predictive-processing framework, repeated standards become increasingly expected and may therefore elicit attenuated processing as the regularity of the sequence is learned. The larger N250 response to standards in infants exposed to higher prenatal maternal anxiety is therefore more compatible with reduced attenuation or altered weighting of predictable input than with an enhanced prediction-error response. It may reflect slower habituation or a lower tendency to treat repetitive information as irrelevant. In contrast, the smaller N250 associated with higher maternal mindfulness is compatible with more efficient attenuation or allocation of fewer processing resources to the same frequently repeated input. The larger P150 associated with mindfulness may additionally indicate differences in an earlier stage of auditory attention, although its functional interpretation should remain cautious.

The absence of anxiety-related effects for any of the three deviant types argues against a generalized enhancement of neural responses to unexpected events and points more specifically to differences in the processing of highly repetitive, predictable input. The opposite associations of anxiety and mindfulness with the standard-tone response are consistent with a broader differential-calibration principle: both positive maternal psychological characteristics and maternal distress may be associated with how the developing brain allocates relevance and processing resources to recurring signals, but in different ways. These ERP findings do not directly measure precision, and habituation, attention, and sensory adaptation remain plausible alternative explanations.

## 4.2 Publication 10: Emotional face-voice processing at 9 months

In this publication, maternal anxiety was assessed before 15 weeks of gestation using both the State Anxiety subscale of the State Trait Anxiety Inventory (STAI) and the anxiety subscale of the Symptom Checklist-90. Both were analysed as continuous predictors in repeated-measures ANCOVAs, but significant findings emerged only for the broader anxiety symptoms measured with the SCL-90. Relevant maternal, perinatal and infant characteristics, including postnatal maternal anxiety, were considered as potential covariates, and Greenhouse–Geisser adjustment was applied where required.

Infants first viewed a happy or fearful facial expression, which served as the visual prime, and subsequently heard a happy or fearful non-verbal vocalization, creating emotionally congruent and incongruent face-voice combinations (Otte et al., 2015; see Table 4). Each 1,400-ms trial began with presentation of the face for 900 ms; the vocalization was then presented for 500 ms while the face remained visible. The happy or fearful face could thus provide affective context for processing the subsequent vocalization.

Higher prenatal maternal anxiety was associated with larger P350 amplitudes to fearful vocalizations and smaller P350 amplitudes to happy vocalizations, irrespective of whether the preceding visual prime was a happy or fearful face. A similar association with larger P150 amplitudes to fearful vocalizations approached significance. Both associations were robust to adjustment for several potential confounders; the P350 effect was attenuated to a trend, however, after specifically adjusting for postnatal maternal anxiety ($p$ increasing from .043 to .052).

The larger P350 response was interpreted as indicating greater allocation of stimulus-driven attention to fearful vocalizations, whereas the P150 trend may reflect increased early arousal or feature processing. The publication did not provide evidence that prenatal maternal anxiety altered the integration of emotional information across face and voice, because the associations did not depend on whether the facial expression and vocalization conveyed the same emotion.

Within a predictive-processing framework, the opposite P350 associations with fearful and happy vocalizations are compatible with greater weighting of fear-related auditory information relative to positive information, rather than a general increase in emotional processing. Because this pattern was independent of the visual prime's emotion, however, it does not demonstrate altered cross-modal use of context to generate predictions about the vocalization; enhanced exogenous attention or arousal to fearful sounds remains a plausible alternative.

## 4.3 Publication 11: Affective picture processing at 4 years

In the affective-picture publication, maternal anxiety symptoms were assessed during the second trimester using the anxiety subscale of the Symptom Checklist-90 and analysed as a continuous predictor in multiple regression analyses. Gestational age at birth and maternal postnatal anxiety were included as covariates, and false-discovery-rate correction was applied for multiple comparisons.

Children viewed neutral, pleasant, and unpleasant pictures selected from the International Affective Picture System while the late positive potential (LPP) was recorded (Lang et al., 2008; see Table 4). The LPP is commonly interpreted as an index of sustained attentional and motivational processing of affective stimuli (Hajcak et al., 2010).

Contrary to the hypothesis that prenatal maternal anxiety would be associated with stronger processing of unpleasant pictures, no such association was found. Instead, higher maternal anxiety was associated with larger middle and late LPP amplitudes to neutral pictures at both central and anterior electrode sites. These associations remained significant after adjustment for maternal postnatal anxiety and gestational age at birth; after false-discovery-rate correction, the late-window associations remained significant while the middle-window associations became only marginally significant (van den Heuvel et al., 2018b).

The inclusion of pleasant and unpleasant pictures is important for interpreting the specificity of this finding. Prenatal maternal anxiety was not associated with a generalized increase in sustained processing across affective categories, nor with a straightforward negativity bias. Rather, the effect was confined to neutral stimuli. Within a predictive-processing framework, neutral pictures may carry greater uncertainty about their affective or motivational significance than clearly pleasant or unpleasant pictures. The larger LPP may therefore reflect more sustained evaluation of stimuli whose relevance is not immediately resolved.

One possible interpretation is that children prenatally exposed to higher maternal anxiety showed enhanced vigilance for potential threat in ambiguous or apparently neutral input, adaptive in uncertain contexts but costly if neutral information is repeatedly treated as significant. This account predicts that directly manipulating ambiguity or threat expectancy, not attempted here, should reproduce the effect. An alternative, non-PP explanation, that neutral pictures are simply less perceptually resolved or less memorable than clearly valenced images regardless of any anxiety-related weighting, cannot be ruled out by this design and would need to be addressed by manipulating image ambiguity directly in future work. The pleasant-picture condition serves as a comparison showing that the association was specific to neutral rather than generally positive or negative material, and does not itself provide evidence for beneficial effects of positive prenatal experience.

## 4.4 Predictive-processing interpretation of the infant and child findings

Across the three Tilburg publications, maternal psychological characteristics during the first half of pregnancy were associated with selective differences in infant and child processing. Higher maternal anxiety was associated with increased processing of a highly repetitive standard tone, stronger processing of fearful relative to happy vocalizations, and greater sustained attention to neutral rather than pleasant or unpleasant pictures. In the auditory oddball study, maternal mindfulness showed an association in the opposite direction to anxiety for the same repeated standard stimulus.

Within a PP framework, the findings are compatible with differences in how recurring, affectively salient, and uncertain sensory information is assigned relevance and processing resources: less attenuation of highly predictable input, greater weighting of fear-related auditory information, and prolonged evaluation of neutral information whose significance may be less immediately resolved. The face-voice result did not, however, show altered use of the visual prime to generate cross-modal affective predictions.

The comparison conditions constrain simpler accounts. No anxiety-related association was detected for the three auditory deviant types; the face-voice study provided no robust evidence that the anxiety association depended on visual-prime congruency; and no association was found for unpleasant pictures, with pleasant-picture associations also non-significant. The pattern therefore does not reduce to generalized deviance amplification or a simple negativity bias. Sensory attenuation, habituation, arousal, and attentional allocation remain viable non-PP explanations, and are considered further in Section 6.

### 4.5 Relation to developmental psychopathology in the Tilburg cohort

The ERP publications were not designed to assess offspring psychopathology directly. Findings from the same longitudinal cohort nevertheless indicate that prenatal maternal anxiety was associated with later socio-emotional and behavioural development, often in interaction with the postnatal caregiving environment. Maternal anxiety during pregnancy mediated the association between maternal mindfulness and infant self-regulation (van den Heuvel, Johannes, et al., 2015). The association between prenatal maternal anxiety and child internalizing problems in early childhood was accounted for by continuity in concurrent maternal anxiety and, subsequently, mindful parenting (Henrichs et al., 2021). In a study examining frontal alpha asymmetry and negative affect, a more complex environmental-sensitivity pattern emerged for externalizing behaviour at age 4: children with both high negative affect and greater left-sided frontal alpha asymmetry showed the most externalizing behaviour when mindful parenting was low but the least when mindful parenting was high (Chhangur et al., 2025).

These findings indicate that developmental outcomes in the Tilburg cohort cannot be understood from prenatal exposure alone. They highlight both continuity in maternal anxiety and the potential modifying role of positive postnatal caregiving characteristics. Within the present empirical reappraisal, they provide behavioural and developmental context for the early ERP findings and support the broader view that prenatal and postnatal influences may shape outcomes through interacting processes. They do not, however, establish that the neural-processing differences observed in infancy and early childhood caused later socio-emotional or behavioural outcomes.

## 5. A Falsifiable Cross-Cohort Bridging Hypothesis: Developmental Integration and Possible Pathways to Later Vulnerability

The Leuven and Tilburg cohorts offer complementary rather than directly replicating evidence. Leuven identified selective differences in internally organized cognitive control and its neural correlates from adolescence into adulthood, whereas the Tilburg cohort examined earlier sensory, affective, and attentional processing during infancy and early childhood. Read in developmental order, the Tilburg cohort shows differences in responses to repeated, fearful, and neutral stimuli, while Leuven shows later differences when task goals, expected outcomes, feedback, contextual information, and bodily signals must be integrated to guide behaviour.

The developmental reading offered here is a cross-cohort bridging hypothesis, not developmental continuity demonstrated by the data: the Tilburg infants were not followed into the Leuven adolescent and adult outcomes, and no existing cohort combines both sets of measures. Early differences in sensory and affective weighting could plausibly contribute to the calibration of broader regulatory systems, with related differences later expressed less in early ERP amplitudes and more in the internal maintenance of task goals, the integration of competing information, the adaptation of response policies, or the affective evaluation of bodily challenge. To keep this reading falsifiable rather than narrative, we follow the logic of strong inference (Platt, 1964; Bowman et al., 2023) and state it as four propositions, with the direction of each predicted effect specified a priori.

The account predicts one directional signature across ages: weaker context-appropriate precision assignment. Early in development this should surface as *increased* processing of highly predictable input (reduced attenuation of repeated standards, as observed at 9 months); in adolescence and adulthood it should surface as *reduced* endogenous up-weighting of internally generated goals and expected outcomes (reduced endogenous-versus-exogenous prefrontal modulation, as observed at ages 17–20). The surface directions differ, but both follow from the same predicted failure to assign precision according to current context, and the mapping is fixed in advance rather than chosen to fit each result. The account would be contradicted by *attenuated* standard-tone responses in exposed infants, or by *enhanced* endogenous modulation in exposed adults; either observation is a legitimate outcome of the studies proposed below and could not be absorbed post hoc.

**P1 (developmental linkage).** In a cohort assessing both measures, reduced standard-tone attenuation in infancy will predict reduced endogenous-versus-exogenous control modulation in adolescence, over and above concurrent maternal anxiety. As no existing cohort includes both measures, this prediction requires longitudinal assessment of both markers in the same individuals.

**P2 (predictability manipulation).** Direct manipulation of stimulus predictability in prenatal-anxiety cohorts (for example, roving or reversing standards across timescales; Banaschewski et al., 2026; Billing et al., 2025) will yield exposure-related differences for predictable input but not for unexpected input, with prenatal and postnatal maternal anxiety modelled jointly.

**P3 (scaffolding manipulation).** Within-subject removal of external task scaffolding will enlarge exposure-related performance differences, whereas adding explicit external cues will reduce them. Failure of this within-task modulation in an adequately powered sample would count directly against the account.

**P4 (affective-weighting mediation).** Early weighting of threat-related and affectively ambiguous input (fearful-vocalization P350; neutral-picture LPP) will mediate prospective associations between prenatal maternal anxiety and childhood anxiety-proneness, testable in ongoing Tilburg follow-ups.

Prediction errors may arise in several of the paradigms, but the present interpretation concerns primarily the context-dependent weighting and influence of sensory, affective, contextual, feedback-related, and bodily information, rather than differences in prediction-error magnitude itself.

The Tilburg findings may relate more directly to the earlier Leuven observation that higher maternal anxiety at 12–22 weeks was associated with greater self-reported anxiety in offspring at 8–9 years. Greater processing of fearful vocalizations in infancy and neutral pictures at age 4 years may indicate altered weighting of threat-related or affectively uncertain information that could contribute to later anxiety-proneness. This interpretation is consistent with the broader developmental framework proposed by Filippi et al. (2026), who discuss early affective and neural responsivity in relation to later anxiety vulnerability.

Two further differences qualify this cross-cohort reading: Tilburg mothers were on average less anxious than Leuven mothers (see Table 1), and the cohorts analysed anxiety differently, as a high-anxiety group contrast on the STAI-State in Leuven and as a continuous, mostly SCL-90-based predictor in Tilburg. Tests of P1–P4 should therefore use the same anxiety measure and analytic model across ages.

The value of this bridging hypothesis lies in converting the cross-cohort pattern into testable predictions (P1–P4) about relevance, expected outcomes, feedback, context, and bodily signals. PP provides a common vocabulary for describing this possible continuity, but leaves open whether the same computations underlie each finding.

These processes may contribute to later vulnerability in interaction with ongoing maternal anxiety, caregiving quality, and later environmental demands. If a child repeatedly allocates greater processing resources to repetitive, threat-related, or ambiguous input, the environment may remain more demanding and less readily filtered. This could contribute to vigilance, distractibility, difficulty disengaging attention, emotional over-reactivity, avoidance, irritability, sleep problems, or broader self-regulatory difficulties. At later ages, less efficient internally organized control may become particularly relevant when explicit guidance is unavailable or several sources of information must be integrated.

Psychopathology would not be expected to emerge from one isolated perceptual or cognitive difference, but from repeated interactions among neural processing, bodily regulation, caregiving context, and later demands; the broader neurobiological substrates potentially implicated in this account are addressed in the companion theoretical paper (Frasch and Van den Bergh, 2026) rather than tested by these eleven publications.

## 6. Interpretational Constraints and the Falsifiability of the Broader PP Account

Although the findings are compatible with predictive-processing interpretations, the original publications were not designed as formal tests of predictive coding or predictive processing. The term "predictive coding" typically refers to the specific hierarchical, message-passing implementation of predictive processing (Rao and Ballard, 1999), whereas we use “predictive processing” throughout for the broader framework. A first constraint concerns evidential independence, which limits the strength of any cross-publication inference. The two cohorts are independent, but the publications from each cohort are interrelated longitudinal follow-ups rather than independent replications. This interdependence is inseparable from a central strength of the cohorts: repeated assessments across developmental stages and modalities make it possible to examine whether a coherent developmental pattern emerges over time. The resulting convergence should therefore be interpreted as qualitative and theory-guided, rather than as evidence from matched or independent replications. Statistical modelling and correction for multiple comparisons varied across publications and analytical modalities. Some analyses applied formal correction procedures, whereas others used uncorrected significance thresholds. In modest exploratory samples, stricter correction reduces the risk of false-positive findings but may also increase the risk of false-negative findings. This matters especially because several central inferences rely on dissociations: a difference was detected in one condition or outcome but not in a theoretically informative comparison condition or outcome. We refer to these non-significant results in the comparison conditions as key nulls. In the Leuven samples that contributed key nulls, group sizes ranged from 8 versus 9 ($N = 17$) to 16 versus 41 ($N = 57$), corresponding to approximate minimum detectable effects at 80% power ($\alpha = .05$, two-tailed, simple two-group contrast) of about $d = 0.8$ to $d = 1.4$ (see Table 5). The continuous Tilburg analyses ($n = 79$–$86$) could detect bivariate associations of approximately $|r| \geq .30$. These values are only rough benchmarks because the original analyses often included covariates or repeated-measures factors. Supplementary Table S4 therefore audits both detected findings and the null or comparison findings that carry an inferential role in the present synthesis.

The original publications generally reported these key nulls as non-significant tests rather than as equivalence tests. Standardized effect sizes and confidence intervals were not consistently available. Where an exact test statistic was reported, an approximate standardized contrast can sometimes be reconstructed, for example for the Stop-task SSRT result in Van den Bergh et al. (2005), but its confidence interval remains wide; in other reports only upper bounds on F values or qualitative statements of non-significance were provided, precluding a precise effect-size estimate. The Tilburg comparison nulls likewise vary in how directly their uncertainty can be quantified (see Supplementary Table S4). These findings therefore do not establish that the comparison functions were preserved or unaffected. The dissociative claims should instead be read as differences detected for endogenous or focal outcomes but not for corresponding exogenous or comparison conditions, under limited and only approximately comparable statistical power.

Among the Leuven Tier 1 positive group-contrast findings that reached significance, observed effect sizes ranged from about $d = 0.7$–$1.0$ in the publications with larger full samples (Van den Bergh et al., 2005, 2006; Mennes et al., 2006; $N = 49$–$64$; the 2006 effect was confined to boys, see Supplementary Table S4) to about $d = 0.97$–$2.5$ in the smallest samples (Mennes et al., 2009, 2020; $N = 17$–$23$), with correspondingly wide confidence intervals. This inverse relationship between sample size and observed effect size is compatible with genuinely larger effects in the more demanding endogenous-control contrasts, but underpowered studies that reach significance also tend to overestimate the true effect size (Ioannidis, 2008; Button et al., 2013). The exact magnitude of the smallest-sample effects should therefore be treated cautiously, even though the reported contrasts were statistically significant.

Table 5. Evidence audit: design, exposure, and power for the eleven reappraised publications.

| Publication | Cohort, age | Analysed N | Exposure operationalization | MDE (80% power) | Covariate robustness (as reported) |
|---|---|---|---|---|---|
| Van den Bergh et al. (2005) | Leuven, 14–15 y | 57 (16 HA / 41 LMA) | STAI-State ≥ 43 at 12–22 wk (dichotomous) | d ≈ 0.83 | Adjusted for other pregnancy periods + postnatal anxiety; RT effect survived IQ check post hoc, error effect attenuated |
| Van den Bergh et al. (2006) | Leuven, 14–15 y | 64 (16 HA / 48 LMA); boys only: 11 HA / 23 LMA | STAI-State > 43 (sample P75 = 43.5) at 12–22 wk (dichotomous) | d ≈ 0.81 (full sample, 16 HA/48 LMA); d ≈ 1.03 (boys only, 11 HA/23 LMA) | Adjusted for other pregnancy periods, postnatal anxiety, parental education, age; effect survived IQ check post hoc |
| Mennes et al. (2006) | Leuven, 17 y | 49 (15 HA / 34 LMA) | STAI-State ≥ 43 at 12–22 wk (dichotomous) | d ≈ 0.87 | Adjusted for other pregnancy periods, postnatal anxiety, non-verbal IQ |
| Mennes et al. (2009) | Leuven, 17 y (boys) | 23 (8 HA / 15 LMA) | STAI-State ≥ 43 at 12–22 wk (dichotomous) | d ≈ 1.23 | Adjusted for other pregnancy periods + postnatal anxiety |
| Mennes et al. (2020) | Leuven, 20 y (men) | 17 (8 HA / 9 LMA) | STAI-State ≥ 43 at 12–22 wk (dichotomous); extreme groups matched on Performance IQ | d ≈ 1.36 | Other pregnancy periods, birth weight, parental education, postnatal anxiety checked (ns) |
| Van den Bergh et al. (2025) | Leuven, 28 y | 48 primary (12 HA / 36 LMA) | STAI-State ≥ 43 at 12–22 wk (dichotomous) | d ≈ 0.93 | Adjusted for birth weight + postnatal anxiety + scanner upgrade; maternal/paternal age, BMI, cigarettes, caffeine, alcohol, offspring education, and parental education/SES all tested for group balance* (all ns, none entered as covariate) |
| Turk et al. (2023) | Leuven, 28–29 y | 49 (13 HA / 36 LMA) | STAI-State ≥ 43 at 12–22 wk (dichotomous) | d ≈ 0.91 | Adjusted for sex, birth weight, postnatal anxiety; maternal/paternal age, BMI, |

| Publication | Cohort, age | Analysed N | Exposure operationalization | MDE (80% power) | Covariate robustness (as reported) |
|---|---|---|---|---|---|
| | | | | | cigarettes, caffeine, alcohol, offspring education, social class, and parental education all tested for group balance* (all ns, none entered as covariate) |
| von Leupoldt et al. (2017) | Leuven, 28 y | 40 (10 HA / 30 LMA) | STAI-**Trait**, 12–22 wk (dichotomous; deviates from corpus) | d ≈ 1.02 | Duration of pregnancy, birthweight, maternal smoking, caffeine, and alcohol use during pregnancy checked (all ns); no concurrent adult state or trait anxiety difference |
| van den Heuvel, Donkers, et al. (2015) | Tilburg, 9 mo | 79 | SCL-90 anxiety and FMI mindfulness, continuous, 2nd trimester (~21 wk) | \|r\| ≈ .31 | GA, birth weight, postnatal anxiety checked (none entered, ns) |
| Otte et al. (2015) | Tilburg, 9 mo | 82 | SCL-90 anxiety continuous, 1st trimester (<15 wk) (STAI-State ns) | \|r\| ≈ .31 | P350 effect attenuated to trend (p = .052) after postnatal-anxiety adjustment |
| van den Heuvel et al. (2018b) | Tilburg, 4 y | 86 | SCL-90 anxiety continuous, 2nd trimester | \|r\| ≈ .30 | Adjusted for GA + postnatal anxiety; late windows survived FDR, middle windows marginal |

*Note. MDE = minimum detectable effect at 80% power (α = .05, two-tailed) for a simple two-group contrast (Cohen's d) or bivariate association (r); MDEs do not account for covariates or repeated-measures structure and are therefore approximate lower bounds on detectable effects. *A covariate that does not differ between groups cannot, by design, account for a between-group difference in the outcome (Miller and Chapman, 2001); where the original publications tested additional demographic or lifestyle variables for group balance, these are listed; none differed significantly between groups.* The full evidence audit, with effect sizes (95% CI) and evidential-weight tier for each finding, is provided in Supplementary Table S4.

Response and retention rates also differed between cohorts: the Leuven cohort maintained response rates of at least 60% through the age-28 assessments, whereas retention in the Tilburg cohort was lower; the resulting infant and child samples were nonetheless comparable to or larger than typical developmental EEG studies (median N = 51; Morales et al., 2025), and comparable in size to other prenatal-distress neuroimaging cohorts.

A second constraint concerns the specificity of PP constructs. None of the original publications formally estimated computational precision, learning rates, uncertainty, volatility, or directional prediction-error signalling. ERP amplitudes, connectivity, white-matter indices, task performance, and dyspnoea unpleasantness may reflect overlapping processes, including attention, arousal, sensory sensitivity, habituation, appraisal, task engagement, and developmental maturation. Predictive-coding interpretations of auditory regularity also depend on unresolved questions about what constitutes a prediction, which generative model produces it, and how neural responses to repetition and change relate to perceptual organization (Denham and Winkler, 2020). PP-based interpretations should therefore specify the proposed mechanism (for example, altered precision weighting rather than another form of prediction-error reduction) and the findings that would support or challenge it (Bowman et al., 2023; Kwisthout et al., 2017). More broadly, the Bayesian-brain framework has been criticized for moving ambiguously between metaphor, formal model, and biologically plausible mechanism, and for accommodating diverse findings post hoc (Mangalam, 2025). PP is most useful here when it generates a precise and testable account rather than simply relabelling observed processes (Hodson et al., 2024). Instrument sensitivity may also have differed across measures: in the Tilburg cohort, the SCL-90 anxiety subscale detected associations that the STAI-State scale did not (Section 4.2), possibly because the SCL-90 weighs somatic symptoms (e.g., trembling, restlessness) more heavily than the STAI-State, which focuses more on psychological symptoms such as worry and fearful thoughts (Bech, 2011); somatic anxiety symptoms may plausibly have a stronger association with fetal physiological and neurodevelopmental processes than psychological symptoms, a pattern we previously noted for a comparable SCL-90-specific effect on maternal cortisol (van den Heuvel et al., 2018a). We cannot exclude, however, that the different patterns of results partly reflect sampling error, because absolute anxiety levels in this non-clinical sample were also lower than in Leuven and closer to a measurement floor for the STAI-State.

A third constraint concerns exposure. Both cohorts assessed variation in maternal anxiety during pregnancy within non-clinical community samples (see Supplementary Tables S2 and S3). They did not primarily examine severe, uncontrollable, or unpredictable adversity, which is central to broader accounts of early-life adversity and environmental unpredictability (Davis and Glynn, 2024; Ugarte and Hastings, 2024). As stated in the Introduction, the findings therefore cannot be used to evaluate claims that traumatic adversity produces explicit threat-related or pessimistic priors, as proposed in some PP accounts of psychological trauma and post-traumatic stress disorder (Linson et al., 2020; Wilkinson et al., 2017). They are most directly relevant to the more parsimonious hypothesis that maternal anxiety during pregnancy is associated with differences in the allocation of relevance and processing resources and in internally organized regulation.

Building on the study-specific PP interpretations in Sections 3.1–4.4, the falsifiability debate is useful because it requires PP accounts to specify which empirical patterns would support them and which would count against them, rather than accommodating any result retrospectively (Bowman et al., 2023). Section 5 applies this logic prospectively, stating the cross-cohort bridging hypothesis as four a priori predictions (P1–P4); the constraints considered here instead concern what the eleven publications, as already published, do and do not establish. The present findings do not confirm or falsify PP as a general framework, but they limit the interpretations that can plausibly be derived from it.

Within the power and measurement limits described above, the selectivity patterns summarized in Sections 3.5 and 4.4 constrain generalized alternatives. In Leuven, they limit generalized cognitive-control and later generalized-deficit accounts; in Tilburg, they limit accounts based on generalized auditory-deviance amplification, visual-prime-driven cross-modal prediction, or a simple negativity bias.

Thus, the value of the reappraised findings from a PP perspective lies in distinguishing among competing PP-compatible explanations rather than in demonstrating the framework itself.

Future research should test these interpretations more directly by manipulating predictability, uncertainty, volatility, and sensory reliability; estimating trial-by-trial learning and precision parameters; varying endogenous and externally cued control demands; and examining effective connectivity or directional message passing. In infants, experimentally manipulating expected and unexpected input while controlling for attention, as in Billing et al. (2025), who examined concurrent (postnatal) maternal anxiety, provides one example of a more direct operationalization of prediction-error processing. Applied prospectively to prenatal cohorts, such paradigms should assess prenatal and postnatal maternal anxiety jointly to distinguish antenatal from concurrent environmental influences. Auditory paradigms could additionally reverse standards and deviants over multiple timescales and combine event-related potentials with dynamic causal modelling to distinguish adaptation, intrinsic gain changes, and descending predictive influences, as illustrated by Banaschewski et al. (2026). Prospective cohorts should assess prenatal and postnatal influences, supportive as well as adverse conditions, sex-specific effects, and repeated outcomes from infancy into adulthood. The present PP interpretation would be weakened if direct manipulations of expectedness or uncertainty failed to reveal the proposed differences, if non-PP attentional models consistently accounted for the data better, or if the distinction between endogenous and externally cued control did not replicate in adequately powered samples.

## 7. Conclusion

Across eleven interrelated publications from two longitudinal prenatal-anxiety cohorts, variation in maternal anxiety during pregnancy was associated with selective differences in sensory, affective, attentional, cognitive, neural, and interoceptive-affective processing. In the Leuven cohort, working memory, visual orienting, and externally cued response inhibition showed no detectable differences, whereas differences recurred when cognitive control had to be generated and maintained endogenously. These findings were accompanied by ERP and task-related fMRI differences and, in adulthood, by selective white-matter, prefrontal-connectivity, Vocabulary, and interoceptive-affective findings. In the Tilburg cohort, maternal anxiety during pregnancy was associated with increased processing of repeated standard tones, fearful relative to happy vocalizations, and neutral rather than pleasant or unpleasant pictures. Maternal mindfulness showed an association in the opposite direction to anxiety for the repeated standard, suggesting that positive and distress-related prenatal states may influence partly overlapping early processing mechanisms in different ways.

Taken together, the findings are compatible with a parsimonious PP-informed account in which maternal anxiety during pregnancy is associated with differences in how relevance and processing resources are allocated to predictable, threat-related, affectively ambiguous, contextual, feedback-related, and bodily information, and in how regulation becomes internally organized. They do not require assumptions about explicit negative beliefs or entrenched pessimistic priors.

This empirical reappraisal suggests that findings from the two longitudinal cohorts form a selective, qualitatively coherent cross-modal pattern that PP can organize more precisely than generalized accounts of impairment or heightened reactivity. Read alongside the companion theoretical paper, it provides a focused developmental hypothesis for direct experimental and computational evaluation.

## Declarations

Data availability. This article reinterprets previously published data from the Tilburg Prenatal Early Life Stress cohort and the Leuven Prenatal Project; no new data were generated.

Funding. The PELS study was supported by the national funding agencies participating in the European Science Foundation EUROCORES Programme EuroSTRESS (EuroSTRESS-PELS, grant 99930AB6-0CAC-423B-9527-7487B33085F3), including the Brain and Cognition Programme of the Netherlands Organisation for Scientific Research (NWO). Both the PELS and Leuven cohorts were partly supported by the European Commission Seventh Framework Programme (FP7-HEALTH-2011.2.2.2-2, BRAINAGE, grant agreement 279281). The Leuven cohort was additionally supported by the Research Foundation - Flanders (FWO-Vlaanderen; projects 1197285N and 1197287N). BRHVdB was further supported by FWO-Vlaanderen SBO project S003524N and by COST Action CA22114, funded by the European Cooperation in Science and Technology.

Competing interests. The authors declare no competing interests.

## Author contributions

B.R.H.V.d.B. conceptualized the empirical synthesis, selected and synthesized the findings across the eleven publications, and developed their initial predictive-processing interpretations. M.G.F. critically evaluated these interpretations and contributed to their theoretical refinement and alignment with the companion framework. B.R.H.V.d.B. drafted the manuscript. Both authors critically revised the manuscript and approved the final version.

## Acknowledgements

We gratefully acknowledge all offspring and their parents who participated in the Leuven Prenatal Project and the Tilburg Prenatal Early Life Stress cohort across the different phases of these longitudinal studies. Their sustained commitment made the research summarized in this article possible. We also thank all co-authors of the original publications for their essential contributions to the design, data collection, analysis, interpretation, and reporting of the individual studies.

## Declaration of generative AI and AI-assisted technologies in the manuscript preparation process

During the preparation of this work, the authors used ChatGPT and Claude to assist with literature organization, drafting, and language editing. The authors reviewed and edited the content as needed and take full responsibility for the content of the article.

## References

Adams, R.A., Stephan, K.E., Brown, H.R., Frith, C.D., Friston, K.J., 2013. The computational anatomy of psychosis. Frontiers in Psychiatry, 4, 47. https://doi.org/10.3389/fpsyt.2013.00047

Andrews-Hanna, J.R., Smallwood, J., Spreng, R.N., 2014. The default network and self-generated thought: Component processes, dynamic control, and clinical relevance. Annals of the New York Academy of Sciences, 1316(1), 29–52. https://doi.org/10.1111/nyas.12360

Arthur, T., Vine, S.J., Buckingham, G., Brosnan, M., Wilson, M.R., Harris, D.J., 2023. Testing predictive coding theories of autism spectrum disorder using models of active inference. PLOS Computational Biology, 19(9), e1011473. https://doi.org/10.1371/journal.pcbi.1011473

Banaschewski, M.T., Mathys, C., Winkler, I., Todd, J., Auksztulewicz, R., 2026. Predictive Processing Over the Course of Aging: Multiple Timescales of Effective Connectivity. European Journal of Neuroscience, 63(1), e70387. https://doi.org/10.1111/ejn.70387

Bandettini, A., Giofrè, D., Panesi, S., Morra, S., Traverso, L., 2025. Latent structure of executive function in preschoolers: A systematic review and meta-analysis. Cognitive Development, 75, 101612. https://doi.org/10.1016/j.cogdev.2025.101612

Barrett, L.F., Simmons, W.K., 2015. Interoceptive predictions in the brain. Nature Reviews Neuroscience, 16(7), 419-429. https://doi.org/10.1038/nrn3950

Bech, P., 2011. Measuring states of anxiety with clinician-rated and patient-rated scales. In: Selek, S. (Ed.), Different Views of Anxiety Disorders. InTech, Croatia, pp. 169–184.

Berger, A., Posner, M.I., 2023. Beyond Infant's Looking: The Neural Basis for Infant Prediction Errors. Perspectives on Psychological Science, 18(3), 664–674. https://doi.org/10.1177/17456916221112918

Billing, A.D.N., Smith, E.S., Cooper, R.J., Lawson, R.P., 2025. Maternal anxiety shapes prediction error responses in the infant brain. Neurophotonics, 12(3), 035013. https://doi.org/10.1117/1.NPh.12.3.035013

Bowman, H., Collins, D.J., Nayak, A.K., Cruse, D., 2023. Is predictive coding falsifiable? Neuroscience and Biobehavioral Reviews, 154, 105404. https://doi.org/10.1016/j.neubiorev.2023.105404

Braeken, M.A.K.A., Kemp, A.H., Outhred, T., Otte, R.A., Monsieur, G.J.Y., Jones, A., Van den Bergh, B.R.H., Van den Bergh, O., 2013. Pregnant mothers with resolved anxiety disorders and their offspring have reduced heart rate variability: Implications for the health of children. PLOS ONE, 8(12), e83186. https://doi.org/10.1371/journal.pone.0083186

Button, K.S., Ioannidis, J.P.A., Mokrysz, C., Nosek, B.A., Flint, J., Robinson, E.S.J., Munafò, M.R., 2013. Power failure: why small sample size undermines the reliability of neuroscience. Nature Reviews Neuroscience, 14(5), 365–376. https://doi.org/10.1038/nrn3475

Chhangur, R.R., Van den Bergh, B.R.H., Hillekens, J., van den Heuvel, M.I., 2025. How negative affect moderates the effect of mindful parenting on child externalizing behavior: Frontal alpha asymmetry as environmental sensitivity factor. Development and Psychopathology. Advance online publication. https://doi.org/10.1017/S0954579425100291

Clark, A., 2013. Whatever next? Predictive brains, situated agents, and the future of cognitive science. Behavioral and Brain Sciences, 36(3), 181–204. https://doi.org/10.1017/S0140525X12000477

Costa, C., Pezzetta, R., Masina, F., Lago, S., Gastaldon, S., Frangi, C., Genon, S., Arcara, G., Scarpazza, C., 2024. Comprehensive investigation of predictive processing: A cross- and within-cognitive domains fMRI meta-analytic approach. Human Brain Mapping, 45(12), e26817. https://doi.org/10.1002/hbm.26817

Davis, E.P., Glynn, L.M., 2024. Annual Research Review: The power of predictability—patterns of signals in early life shape neurodevelopment and mental health trajectories. Journal of Child Psychology and Psychiatry, 65(4), 508–534. https://doi.org/10.1111/jcpp.13958

Denham, S.L., Winkler, I., 2020. Predictive coding in auditory perception: Challenges and unresolved questions. European Journal of Neuroscience, 51(5), 1151–1160. https://doi.org/10.1111/ejn.13802

De Asis-Cruz, J., Krishnamurthy, D., Zhao, L., Kapse, K., Vezina, G., Andescavage, N., Quistorff, J., Lopez, C., Limperopoulos, C., 2020. Association of prenatal maternal anxiety with fetal regional brain connectivity. JAMA Network Open, 3(12), e2022349. https://doi.org/10.1001/jamanetworkopen.2020.22349

Dean, D.C., III, Planalp, E.M., Wooten, W., Kecskemeti, S.R., Adluru, N., Schmidt, C.K., Frye, C., Birn, R.M., Burghy, C.A., Schmidt, N.L., Styner, M.A., Short, S.J., Kalin, N.H., Goldsmith, H.H., Alexander, A.L., Davidson, R.J., 2018. Association of prenatal maternal depression and anxiety symptoms with infant white matter microstructure. JAMA Pediatrics, 172(10), 973–981. https://doi.org/10.1001/jamapediatrics.2018.2132

Filippi, C.A., Massera, A., Xing, J., Martinez Agulleiro, L., 2026. Early-life neural correlates of behavioral inhibition and anxiety risk. Neuropsychopharmacology, 51(1), 95–113. https://doi.org/10.1038/s41386-025-02235-8

Fletcher, P.C., Frith, C.D., 2009. Perceiving is believing: A Bayesian approach to explaining the positive symptoms of schizophrenia. Nature Reviews Neuroscience, 10(1), 48–58. https://doi.org/10.1038/nrn2536

Frasch, M.G., Van den Bergh, B.R.H., 2026. Maladaptive precision: A channel-specific predictive-processing framework for the neurodevelopmental consequences of early life adversity. PsyArXiv preprint (under revision at Neuroscience and Biobehavioral Reviews). https://doi.org/10.31234/osf.io/zj4bf_v1

Friston, K.J., 2010. The free-energy principle: A unified brain theory? Nature Reviews Neuroscience, 11(2), 127–138. https://doi.org/10.1038/nrn2787

Friston, K.J., 2018. Does predictive coding have a future? Nature Neuroscience, 21(8), 1019–1021. https://doi.org/10.1038/s41593-018-0200-7

Friston, K.J., Stephan, K.E., Montague, R., Dolan, R.J., 2014. Computational psychiatry: The brain as a phantastic organ. The Lancet Psychiatry, 1(2), 148–158. https://doi.org/10.1016/S2215-0366(14)70275-5

Graham, R.M., Jiang, L., McCorkle, G., Bellando, B.J., Sorensen, S.T., Glasier, C.M., Ramakrishnaiah, R.H., Rowell, A.C., Coker, J.L., Ou, X., 2020. Maternal anxiety and depression during late pregnancy and newborn brain white matter development. American Journal of Neuroradiology, 41(10), 1908–1915. https://doi.org/10.3174/ajnr.A6759

Hajcak, G., MacNamara, A., Olvet, D.M., 2010. Event-related potentials, emotion, and emotion regulation: An integrative review. Developmental Neuropsychology, 35(2), 129–155. https://doi.org/10.1080/87565640903526504

Harrison, O.K., Köchli, L., Marino, S., Luechinger, R., Hennel, F., Brand, K., Hess, A.J., Frässle, S., Iglesias, S., Vinckier, F., Petzschner, F.H., Harrison, S.J., Stephan, K.E., 2021. Interoception of breathing and its relationship with anxiety. Neuron, 109(24), 4080–4093.e8. https://doi.org/10.1016/j.neuron.2021.09.045

Harvison, K.W., Molfese, D.L., Woodruff-Borden, J., Weigel, R.A., 2009. Neonatal auditory evoked responses are related to perinatal maternal anxiety. Brain and Cognition, 71(3), 369–374. https://doi.org/10.1016/j.bandc.2009.06.004

Hauser, T.U., Will, G.-J., Dubois, M., Dolan, R.J., 2019. Annual Research Review: Developmental computational psychiatry. Journal of Child Psychology and Psychiatry, 60(4), 412–426. https://doi.org/10.1111/jcpp.12964

Gijsen, S., Grundei, M., Blankenburg, F., 2022. Active inference and the two-step task. Scientific Reports, 12, 17682. https://doi.org/10.1038/s41598-022-21766-4

Hennessey, E.-M.P., Swales, D.A., Markant, J., Hoffman, M.C., Hankin, B.L., Davis, E.P., 2024. Maternal anxiety during pregnancy predicts infant attention to affective faces. Journal of Affective Disorders, 344, 104–114. https://doi.org/10.1016/j.jad.2023.09.031

Henrichs, J., van den Heuvel, M.I., Witteveen, A.B., Wilschut, J., Van den Bergh, B.R.H., 2021. Does mindful parenting mediate the association between maternal anxiety during pregnancy and child behavioral/emotional problems? Mindfulness, 12(2), 370–380. https://doi.org/10.1007/s12671-019-01115-9

Hodson, R., Mehta, M., Smith, R., 2024. The empirical status of predictive coding and active inference. Neuroscience and Biobehavioral Reviews, 157, 105473. https://doi.org/10.1016/j.neubiorev.2023.105473

Hunter, S.K., Mendoza, J.H., D'Anna, K., Zerbe, G.O., McCarthy, L., Hoffman, M.C., Freedman, R., Ross, R.G., 2012. Antidepressants may mitigate the effects of prenatal maternal anxiety on infant auditory sensory gating. American Journal of Psychiatry, 169(6), 616–624. https://doi.org/10.1176/appi.ajp.2012.11091365

Ioannidis, J.P.A., 2008. Why most discovered true associations are inflated. Epidemiology, 19(5), 640–648. https://doi.org/10.1097/EDE.0b013e31818131e7

Kim, H., 2014. Involvement of the dorsal and ventral attention networks in oddball stimulus processing: A meta-analysis. Human Brain Mapping, 35(5), 2265–2284. https://doi.org/10.1002/hbm.22326

Kwisthout, J., Bekkering, H., van Rooij, I., 2017. To be precise, the details do not matter: On predictive processing, precision, and level of detail of predictions. Brain and Cognition, 112, 84–91. https://doi.org/10.1016/j.bandc.2016.02.008

Lang, P.J., Bradley, M.M., Cuthbert, B.N., 2008. International Affective Picture System (IAPS): Affective ratings of pictures and instruction manual. University of Florida.

Lawson, R.P., Rees, G., Friston, K.J., 2014. An aberrant precision account of autism. Frontiers in Human Neuroscience, 8, 302. https://doi.org/10.3389/fnhum.2014.00302

Linson, A., Parr, T., Friston, K.J., 2020. Active inference, stressors, and psychological trauma: A neuroethological model of (mal)adaptive explore–exploit dynamics in ecological context. Behavioural Brain Research, 380, 112421. https://doi.org/10.1016/j.bbr.2019.112421

Mandl, S., Alexopoulos, J., Doering, S., Wildner, B., Seidl, R., Bartha-Doering, L., 2024. The effect of prenatal maternal distress on offspring brain development: A systematic review. Early Human Development, 192, 106009. https://doi.org/10.1016/j.earlhumdev.2024.106009

Mangalam, M., 2025. The myth of the Bayesian brain. European Journal of Applied Physiology, 125, 2643–2677. https://doi.org/10.1007/s00421-025-05855-6

Marlow, L.L., Faull, O.K., Finnegan, S.L., Pattinson, K.T.S., 2019. Breathlessness and the brain: the role of expectation. Current Opinion in Supportive and Palliative Care, 13(3), 200–210. https://doi.org/10.1097/SPC.0000000000000441

Menon, V., D'Esposito, M., 2022. The role of PFC networks in cognitive control and executive function. Neuropsychopharmacology, 47(1), 90–103. https://doi.org/10.1038/s41386-021-01152-w

Mennes, M., Stiers, P., Lagae, L., Van den Bergh, B.R.H., 2006. Long-term cognitive sequelae of antenatal maternal anxiety: Involvement of the orbitofrontal cortex. Neuroscience and Biobehavioral Reviews, 30(8), 1078–1086. https://doi.org/10.1016/j.neubiorev.2006.04.003

Mennes, M., Stiers, P., Lagae, L., Van den Bergh, B.R.H., 2020. Antenatal maternal anxiety modulates the BOLD response in 20-year-old men during endogenous cognitive control. Brain Imaging and Behavior, 14(3), 830–846. https://doi.org/10.1007/s11682-018-0027-6

Mennes, M., Van den Bergh, B.R.H., Lagae, L., Stiers, P., 2009. Developmental brain alterations in 17-year-old boys are related to antenatal maternal anxiety. Clinical Neurophysiology, 120(6), 1116–1122. https://doi.org/10.1016/j.clinph.2009.04.003

Miyake, A., Friedman, N.P., Emerson, M.J., Witzki, A.H., Howerter, A., Wager, T.D., 2000. The unity and diversity of executive functions and their contributions to complex "frontal lobe" tasks: A latent variable analysis. Cognitive Psychology, 41(1), 49–100. https://doi.org/10.1006/cogp.1999.0734

Miller, G.A., Chapman, J.P., 2001. Misunderstanding analysis of covariance. Journal of Abnormal Psychology, 110(1), 40–48. https://doi.org/10.1037/0021-843X.110.1.40

Morales, S., Oh, L., Cox, K., Rodriguez-Sanchez, R., Nadaya, G., Buzzell, G.A., Troller-Renfree, S.V., 2025. Generalizability of developmental EEG: Demographic reporting, representation, and sample size. Developmental Cognitive Neuroscience, 74, 101567. https://doi.org/10.1016/j.dcn.2025.101567

Otte, R.A., Donkers, F.C.L., Braeken, M.A.K.A., van den Heuvel, M.I., Winkler, I., Van den Bergh, B.R.H., 2015. Multimodal processing of emotional information in 9-month-old infants II: Prenatal exposure to maternal anxiety. Brain and Cognition, 95, 107–117. https://doi.org/10.1016/j.bandc.2014.12.001

Page, M.J., McKenzie, J.E., Bossuyt, P.M., Boutron, I., Hoffmann, T.C., Mulrow, C.D., et al., 2021. The PRISMA 2020 statement: an updated guideline for reporting systematic reviews. BMJ, 372, n71. https://doi.org/10.1136/bmj.n71

Paulus, M.P., Stein, M.B., 2010. Interoception in anxiety and depression. Brain Structure and Function, 214(5–6), 451–463. https://doi.org/10.1007/s00429-010-0258-9

Platt, J.R., 1964. Strong inference. Science, 146(3642), 347–353. https://doi.org/10.1126/science.146.3642.347

Posner, J., Cha, J., Roy, A.K., Peterson, B.S., Bansal, R., Gustafsson, H.C., Raffanello, E., Gingrich, J., Monk, C., 2016. Alterations in amygdala-prefrontal circuits in infants exposed to prenatal maternal depression. Translational Psychiatry, 6(11), e935. https://doi.org/10.1038/tp.2016.146

Qiu, A., Anh, T.T., Li, Y., Chen, H., Rifkin-Graboi, A., Broekman, B.F.P., Kwek, K., Saw, S.-M., Chong, Y.-S., Gluckman, P.D., Fortier, M.V., Meaney, M.J., 2015. Prenatal maternal depression alters amygdala functional connectivity in 6-month-old infants. Translational Psychiatry, 5(2), e508. https://doi.org/10.1038/tp.2015.3

Rao, R.P.N., Ballard, D.H., 1999. Predictive coding in the visual cortex: A functional interpretation of some extra-classical receptive-field effects. Nature Neuroscience, 2(1), 79–87. https://doi.org/10.1038/4580

Scheinost, D., Sinha, R., Cross, S.N., Kwon, S.H., Sze, G., Constable, R.T., Ment, L.R., 2017. Does prenatal stress alter the developing connectome? Pediatric Research, 81(1–2), 214–226. https://doi.org/10.1038/pr.2016.197

Shipp, S., 2016. Neural elements for predictive coding. Frontiers in Psychology, 7, 1792. https://doi.org/10.3389/fpsyg.2016.01792

Sterzer, P., Adams, R.A., Fletcher, P., Frith, C., Lawrie, S.M., Muckli, L., Petrovic, P., Uhlhaas, P., Voss, M., Corlett, P.R., 2018. The predictive coding account of psychosis. Biological Psychiatry, 84(9), 634–643. https://doi.org/10.1016/j.biopsych.2018.05.015

Turk, E., van den Heuvel, M.I., Sleurs, C., Billiet, T., Uyttebroeck, A., Sunaert, S., Mennes, M., Van den Bergh, B.R.H., 2023. Maternal anxiety during pregnancy is associated with weaker prefrontal functional connectivity in adult offspring. Brain Imaging and Behavior, 17(6), 595–607. https://doi.org/10.1007/s11682-023-00787-1

Ugarte, E., Hastings, P.D., 2024. Assessing unpredictability in caregiver–child relationships: Insights from theoretical and empirical perspectives. Development and Psychopathology, 36(3), 1070–1089. https://doi.org/10.1017/S0954579423000305

Van de Cruys, S., Evers, K., Van der Hallen, R., Van Eylen, L., Boets, B., de-Wit, L., Wagemans, J., 2014. Precise minds in uncertain worlds: predictive coding in autism. Psychological Review, 121(4), 649–675. https://doi.org/10.1037/a0037665

Van de Cruys, S., Van der Hallen, R., Wagemans, J., 2017. Disentangling signal and noise in autism spectrum disorder. Brain and Cognition, 112, 78–83. https://doi.org/10.1016/j.bandc.2016.08.004

Van den Bergh, B.R.H., 1990. The influence of maternal emotions during pregnancy on fetal and neonatal behavior. Pre- and Peri-Natal Psychology Journal, 5(2), 119–130.

Van den Bergh, B.R.H., Antonelli, M.C., Stein, D.J., 2024. Current perspectives on perinatal mental health and neurobehavioral development: focus on regulation, coregulation and self-regulation. Current Opinion in Psychiatry, 37, 237–250. https://doi.org/10.1097/YCO.0000000000000932

Van den Bergh, B.R.H., Marcoen, A., 2004. High antenatal maternal anxiety is related to ADHD symptoms, externalizing problems, and anxiety in 8- and 9-year-olds. Child Development, 75(4), 1085–1097. https://doi.org/10.1111/j.1467-8624.2004.00727.x

Van den Bergh, B.R.H., Mennes, M., Oosterlaan, J., Stevens, V., Stiers, P., Marcoen, A., Lagae, L., 2005. High antenatal maternal anxiety is related to impulsivity during performance on cognitive tasks in 14- and 15-year-olds. Neuroscience and Biobehavioral Reviews, 29(2), 259–269. https://doi.org/10.1016/j.neubiorev.2004.10.010

Van den Bergh, B.R.H., Mennes, M., Stevens, V., van der Meere, J., Börger, N., Stiers, P., Marcoen, A., Lagae, L., 2006. ADHD deficit as measured in adolescent boys with a continuous performance task is related to antenatal maternal anxiety. Pediatric Research, 59(1), 78–82. https://doi.org/10.1203/01.pdr.0000191143.75673.52

Van den Bergh, B.R.H., Mulder, E.J.H., 2012. Fetal sleep organization: A biological precursor of self-regulation in childhood and adolescence? Biological Psychology, 89(3), 584–590.

Van den Bergh, B.R.H., Van Calster, B., Smits, T., Van Huffel, S., Lagae, L., 2008. Antenatal maternal anxiety is related to HPA-axis dysregulation and self-reported depressive symptoms in adolescence: A prospective study on the fetal origins of depressed mood. Neuropsychopharmacology, 33(3), 536–545. https://doi.org/10.1038/sj.npp.1301450

Van den Bergh, B.R.H., Sleurs, C., Geusens, B., Emsell, L., Sunaert, S., Billiet, T., 2025. White matter microstructure and cognitive abilities in 28-year-old offspring prenatally exposed to maternal anxiety: A prospective exploratory multimodal brain imaging study. Brain and Cognition, 188, 106319. https://doi.org/10.1016/j.bandc.2025.106319

Van den Bergh, B.R.H., van den Heuvel, M.I., Lahti, M., Braeken, M., de Rooij, S.R., Entringer, S., Hoyer, D., Roseboom, T., Räikkönen, K., King, S., Schwab, M., 2020. Prenatal developmental origins of behavior and mental health: The influence of maternal stress in pregnancy. Neuroscience and Biobehavioral Reviews, 117, 26–64. https://doi.org/10.1016/j.neubiorev.2017.07.003

van den Heuvel, M.I., Donkers, F.C.L., Winkler, I., Otte, R.A., Van den Bergh, B.R.H., 2015. Maternal mindfulness and anxiety during pregnancy affect infants' neural responses to sounds. Social Cognitive and Affective Neuroscience, 10(3), 453–460. https://doi.org/10.1093/scan/nsu075

van den Heuvel, M.I., van Assen, M.A.L.M., Glover, V., Claes, S., Van den Bergh, B.R.H., 2018a. Associations between maternal psychological distress and salivary cortisol during pregnancy: A mixed-models approach. Psychoneuroendocrinology, 96, 52–60. https://doi.org/10.1016/j.psyneuen.2018.06.005

van den Heuvel, M.I., Henrichs, J., Donkers, F.C.L., Van den Bergh, B.R.H., 2018b. Children prenatally exposed to maternal anxiety devote more attentional resources to neutral pictures. Developmental Science, 21(4), e12612. https://doi.org/10.1111/desc.12612

van den Heuvel, M.I., Johannes, M.A., Henrichs, J., Van den Bergh, B.R.H., 2015. Maternal mindfulness during pregnancy and infant socio-emotional development and temperament: The mediating role of maternal anxiety. Early Human Development, 91(2), 103–108. https://doi.org/10.1016/j.earlhumdev.2014.12.003

Van der Ploeg, H.M., Defares, P.B., Spielberger, C.D., 1980. Handleiding bij de Zelf-Beoordelings Vragenlijst ZBV: Een nederlandstalige bewerking van de Spielberger State-Trait Anxiety Inventory, STAI-DY [Manual of the Self-Evaluation Questionnaire: A Dutch version of the State-Trait Anxiety Inventory]. Lisse, Netherlands: Swets & Zeitlinger.

von Leupoldt, A., Mangelschots, E., Niederstrasser, N.G., Braeken, M., Billiet, T., Van den Bergh, B.R.H., 2017. Prenatal stress exposure is associated with increased dyspnoea perception in adulthood 28 years later. European Respiratory Journal, 50(2), 1700642. https://doi.org/10.1183/13993003.00642-2017

Wilkinson, S., Dodgson, G., Meares, K., 2017. Predictive processing and the varieties of psychological trauma. Frontiers in Psychology, 8, 1840. https://doi.org/10.3389/fpsyg.2017.01840

Wu, Y., De Asis-Cruz, J., Limperopoulos, C., 2024. Brain structural and functional outcomes in the offspring of women experiencing psychological distress during pregnancy. Molecular Psychiatry, 29(7), 2223–2240. https://doi.org/10.1038/s41380-024-02449-0

# Supplementary Material

*Maternal Anxiety During Pregnancy and Offspring Information Processing Across Development: A Theory-Guided Predictive-Processing Reappraisal of Two Longitudinal Cohorts*

Bea R. H. Van den Bergh, Martin G. Frasch

## Figure S1

### PRISMA figure: selection of papers, Leuven and Tilburg cohort

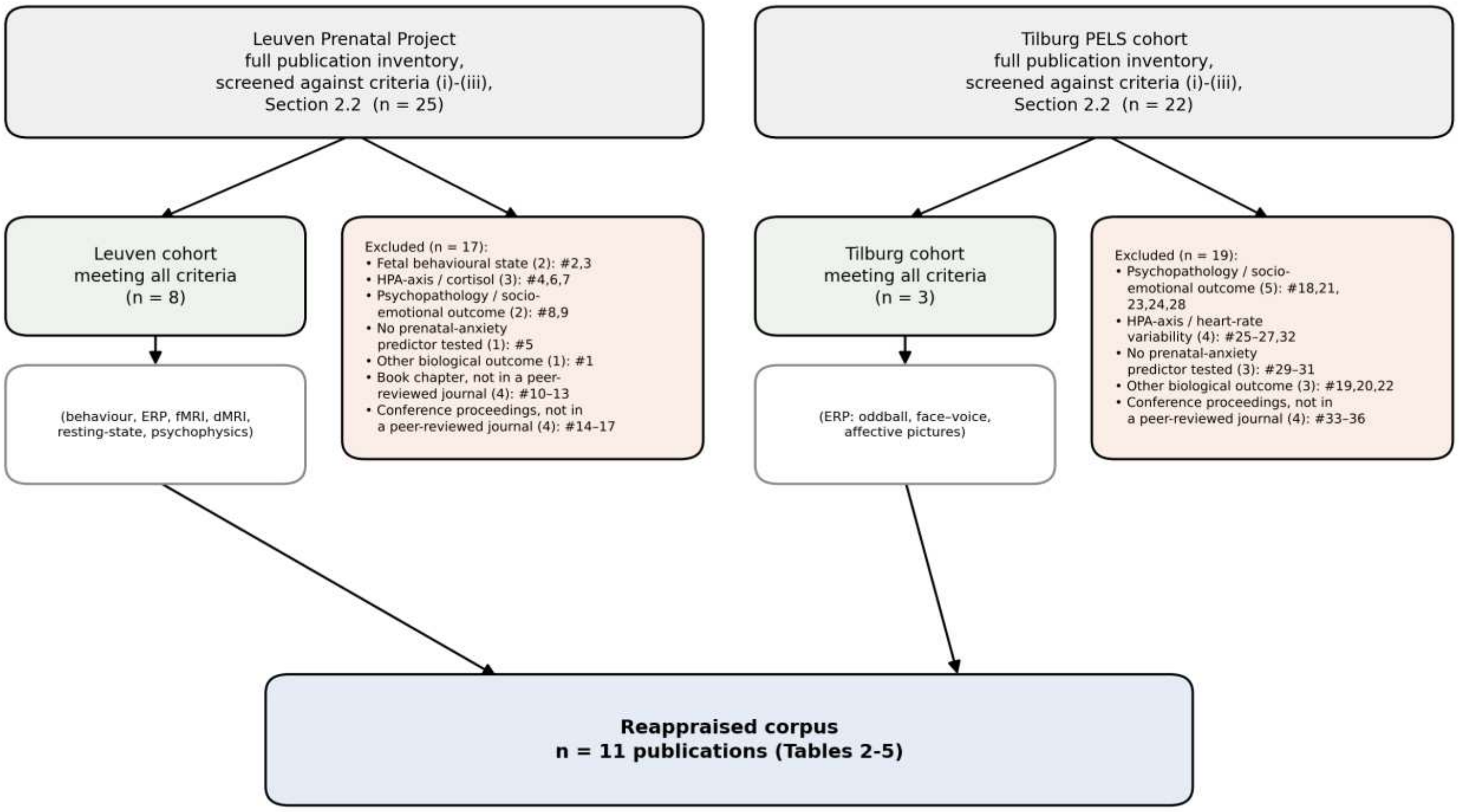

*Excluded publications and their exclusion reasons are listed in full in Supplementary Table S1.*

Note. Selection of publications for the reappraisal. Adapted PRISMA-style flow from the full publication inventories of the two cohorts, through the inclusion criteria (i)–(iii) of Section 2.2, to the eleven reappraised publications (eight Leuven, three Tilburg). Each publication was assigned a single, primary exclusion reason, i.e., the first of the following checks it failed, in this order: publication venue (not published in a peer-reviewed journal, including book chapters and conference proceedings), whether a maternal-anxiety or maternal-mindfulness predictor was tested, and outcome domain (fetal behavioural state; HPA-axis, cortisol or heart-rate variability; psychopathology or socio-emotional outcome; other biological outcome). Numbers after # refer to the entries in Table S1.

# Table S1

## Excluded publications and primary exclusion reason, Leuven and Tilburg cohorts

*Because some publications could have met more than one exclusion criterion, each was assigned to a single, primary exclusion reason following a fixed hierarchical check: publication venue was assessed first (not published in a peer-reviewed journal, including book chapters and conference proceedings), followed by whether a maternal-anxiety or maternal-mindfulness predictor was tested, and finally outcome domain (fetal behavioural state; HPA-axis or heart-rate variability; psychopathology or socio-emotional outcome; or other biological outcome outside the specified information-processing domains). A publication was excluded at the first criterion it failed; subsequent criteria were not evaluated for that publication. Reference numbers correspond to the numbering in this table and are cross-referenced next to each exclusion category in Figure S1.*

*For conference-proceedings entries, the reason column specifies whether the proceedings venue itself implemented formal peer review of the full paper; regardless of this, none of these entries were published in a peer-reviewed journal, so all remain excluded on venue grounds (criterion (iii) of Section 2.2).*

### Leuven cohort (n = 17 excluded)

| No. | Citation | Primary exclusion reason |
|---|---|---|
| 1 | Vos, S., Van den Bergh, B.R.H., Martens, D.S., Bijnens, E., Shkedy, Z., Kindermans, H., Platzer, M., Schwab, M., Nawrot, T.S. (2024). Maternal perceived stress and green spaces during pregnancy are associated with adult offspring gene (NR3C1 and IGF2/H19) methylation patterns in adulthood: A pilot study. Psychoneuroendocrinology, 167, 107088. | Other biological outcome |
| 2 | Van den Bergh, B. (1990). The influence of maternal emotions during pregnancy on fetal and neonatal behavior. Pre- and Peri-Natal Psychology Journal, 5(2), 119-130. | Fetal behavioural state |
| 3 | Van den Bergh, B., Mulder, E.J.H. (2012). Fetal sleep organization: a biological precursor of self-regulation in childhood and adolescence? Biological Psychology, 89(3), 584-590. | Fetal behavioural state |
| 4 | Van den Bergh, B., Van Calster, B. (2009). Diurnal cortisol profiles and evening cortisol in post-pubertal adolescents scoring high on the Children's Depression Inventory. Psychoneuroendocrinology, 34(5), 791-794. | HPA-axis / cortisol |
| 5 | Mennes, M., Wouters, H., Van den Bergh, B., Lagae, L., Stiers, P. (2008). ERP correlates of complex human decision making in a gambling paradigm: detection and resolution of conflict. Psychophysiology, 45(5), 714-720. | No prenatal-anxiety predictor tested |
| 6 | Van den Bergh, B., Van Calster, B., Pinna Puissant, S., Van Huffel, S. (2008). Self-reported symptoms of depressed mood, trait anxiety and aggressive behavior in post-pubertal adolescents: Associations with diurnal cortisol profiles. Hormones and Behavior, 54(2), 253-257. | HPA-axis / cortisol |
| 7 | Van den Bergh, B., Van Calster, B., Smits, T., Van Huffel, S., Lagae, L. (2008). Antenatal maternal anxiety is related to HPA-axis dysregulation and self-reported depressive symptoms in adolescence: a prospective study on the fetal origins of depressed mood. Neuropsychopharmacology, 33(3), 536-545. | HPA-axis / cortisol |
| 8 | Van den Bergh, B., Marcoen, A. (2004). High antenatal maternal anxiety is related to ADHD symptoms, externalizing problems and anxiety in 8/9-year-olds. Child Development, 75, 1085-1097. | Psychopathology / socio-emotional outcome |
| 9 | Van den Bergh, B., Mennes, M. (2006). Het verband tussen angst bij de moeder in de prenatale periode en zelfregulatie in de adolescentie. Kind & Adolescent, 27, 31-43. [Dutch-language secondary report of findings published in Van den Bergh et al., 2005] | Psychopathology / socio-emotional outcome |
| 10 | Van den Bergh, B. (1992). Maternal emotions during pregnancy and fetal and neonatal behaviour. In: J.G. Nijhuis (Ed.), Fetal Behaviour: Developmental and Perinatal Aspects, Ch. 12 (157-178). New York: Oxford University Press. | Venue: book chapter (not peer-reviewed journal) |
| 11 | Van den Bergh, B. (1988). The relationship between maternal emotionality during pregnancy and the behavioral development of the fetus and neonatus. In: P.G. Fedor-Freybergh, M.L.V. Vogel (Eds.), Prenatal and Perinatal Psychology and Medicine (131-142). Camforth: Parthenon. | Venue: book chapter (not peer-reviewed journal) |
| 12 | Van den Bergh, B., Loomans, E.M., Mennes, M. (2015). Early life influences on cognition, behavior, and emotion in humans: From birth to age 20. In: M.C. Antonelli (Ed.), Perinatal Programming of Neurodevelopment, Ch. 15 (315-331). New York: Springer. | Venue: book chapter (not peer-reviewed journal) |
| 13 | Van den Bergh, B. (2004). Stoornis in zelfregulatie bij kinderen en adolescenten: mede een gevolg van foetale programmering? In: L. Goossens, D. Hutsebaut, K. Verschueren (Eds.), Ontwikkeling en levensloop: liber amicorum Alfons Marcoen (55-79). Leuven: Leuven Universitaire Pers. | Venue: book chapter (not peer-reviewed journal) |
| 14 | Van den Bergh, B.R.H. (2012). Prenatal programming of cognition and emotion in humans: From birth to age 20. In: A. Plagemann (Ed.), Perinatal Programming: The State of the Art (199-205). Berlin International Symposium on Perinatal Programming, Berlin, Germany. | Venue: conference proceedings — edited symposium volume; no evidence of formal external peer review (not a peer-reviewed journal) |
| 15 | Van den Bergh, B. (1991). Maternal emotions during pregnancy and fetal and neonatal behavior. In: P. Nijs, B. Leysen, D. Richter (Eds.), Advanced Research in Psychosomatic Obstetrics and Gynaecology (161-170). Leuven: Peeters. | Venue: conference proceedings — edited symposium volume; no evidence of formal external peer review (not a peer-reviewed journal) |
| 16 | Van den Bergh, B., Vandenberghe, K. (1990). Maternal emotionality during pregnancy and fetal behaviour. In: A. Vansteenwegen, A. Van Assche, P. Nijs (Eds.), Human Sexuality and Family Health Sciences (165-169). Leuven: Peeters. | Venue: conference proceedings — edited symposium volume; no evidence of formal external peer review (not a peer-reviewed journal) |
| 17 | Van den Bergh, B., Vandenberghe, K., Daniëls, H., Casaer, P., Marcoen, A. (1989). The relationship between maternal emotionality during pregnancy and the behavioural development of the fetus and neonatus. In: G. Gennser, K. Marsal, N. Svenningen, K. Lindström (Eds.), Fetal and Neonatal Physiological Measurements III (359-363). Malmö: Flenhags Tryckeri. | Venue: conference proceedings — edited symposium volume; no evidence of formal external peer review (not a peer-reviewed journal) |

### Tilburg cohort (n = 19 excluded)

| No. | Citation | Primary exclusion reason |
|---|---|---|
| 18 | de Waal, N., Boekhorst, M.G.B.M., Van den Bergh, B.R.H., van den Heuvel, M.I. (2026). Dispositional Mindfulness and Mindful Parenting in Fathers: Associations with Child Behavioral Problems. Mindfulness, 17(5), 1567-1578. | Psychopathology / socio-emotional outcome |
| 19 | Soerensen, A.E., Van den Bergh, B.R.H., Braeken, M.A.K.A., van den Heuvel, M.I., Martens, D.S., Vos, S., Nawrot, T.S. (2025). Infant buccal telomere length: associations with maternal distress in pregnancy and offspring temperament. Frontiers in Public Health, 13, 1657714. | Other biological outcome |
| 20 | Soerensen, A.E., Vos, S., Braeken, M.A.K.A., van den Heuvel, M.I., Van den Bergh, B.R.H., Nawrot, T.S. (2026). NR3C1 and IGF2/H19 methylation patterns predict infant and preschooler | Other biological outcome |
| | temperament: Links to maternal stress in pregnancy. Psychoneuroendocrinology, 185, 107728. | |
| 21 | Chhangur, R.R., Van den Bergh, B.R.H., Hillekens, J., van den Heuvel, M.I. (2025). How negative affect moderates the effect of mindful parenting on child externalizing behavior: Frontal alpha asymmetry as environmental sensitivity factor. Development and Psychopathology. | Psychopathology / socio-emotional outcome |
| 22 | Cao-Lei, L., van den Heuvel, M.I., Huse, K., Platzer, M., Elgbeili, G., Braeken, M.A.K.A., Otte, R.A., Witte, O.W., Schwab, M., van den Bergh, B.R.H. (2021). Epigenetic Modifications Associated with Maternal Anxiety during Pregnancy and Children's Behavioral Measures. Cells, 10(9), 2421. | Other biological outcome |
| 23 | McLean, M.A., Van den Bergh, B., Baart, M., Vroomen, J., van den Heuvel, M.I. (2020). The latent positive potential (LPP): A neural marker of internalizing problems in early childhood. International Journal of Psychophysiology, 155, 78-86. | Psychopathology / socio-emotional outcome |
| 24 | Henrichs, J., van den Heuvel, M.I., Witteveen, A.B., Wilschut, J., Van den Bergh, B.R.H. (2021). Does Mindful Parenting Mediate the Association between Maternal Anxiety during Pregnancy and Child Behavioral/Emotional Problems? Mindfulness, 12(2), 370-380. | Psychopathology / socio-emotional outcome |
| 25 | van den Heuvel, M.I., van Assen, M.A.L.M., Glover, V., Claes, S., Van den Bergh, B. (2018). Associations between maternal psychological distress and salivary cortisol during pregnancy: A mixed-models approach. Psychoneuroendocrinology, 96, 52-60. | HPA-axis / heart-rate variability |
| 26 | Braeken, M., Jones, A., Otte, R.A., Nyklicek, I., Van den Bergh, B. (2017). Potential benefits of mindfulness during pregnancy on maternal autonomic nervous system function and infant development. Psychophysiology, 54(2), 279-288. | HPA-axis / heart-rate variability |
| 27 | Braeken, M., Jones, A., Widjaja, D., Van Huffel, S., Monsieur, G., Van Oirschot, C.M., Van den Bergh, B. (2015). Anxious women do not show the expected decrease in cardiovascular stress responsiveness as pregnancy advances. Biological Psychology, 111, 83-89. | HPA-axis / heart-rate variability |
| 28 | Van den Heuvel, M., Johannes, M.A., Henrichs, J., Van den Bergh, B. (2015). Maternal mindfulness during pregnancy and infant socio-emotional development and temperament: The mediating role of maternal anxiety. Early Human Development, 91(2), 103-108. | Psychopathology / socio-emotional outcome |
| 29 | Otte, R.A., Donkers, F.C.L., Braeken, M., Van den Bergh, B. (2015). Multimodal processing of emotional information in 9-month-old infants I: Emotional faces and voices. Brain and Cognition, 95, 99-106. | No prenatal-anxiety predictor tested |
| 30 | van den Heuvel, M., Otte, R., Braeken, M., Winkler, I., Kushnerenko, E., Van den Bergh, B. (2015). Differences between human auditory event-related potentials (AERPs) measured at 2 and 4 months after birth. International Journal of Psychophysiology, 97(1), 75-83. | No prenatal-anxiety predictor tested |
| 31 | Otte, R.A., Winkler, I., Braeken, M.A.K.A., Stekelenburg, J.J., van der Stelt, O., Van den Bergh, B. (2013). Detecting violations of temporal regularities in waking and sleeping two-month-old infants. Biological Psychology, 92(2), 315-322. | No prenatal-anxiety predictor tested |
| 32 | Braeken, M., Kemp, A.H., Outhred, T., Otte, R.A., Monsieur, G., Jones, A., Van den Bergh, B. (2013). Pregnant mothers with resolved anxiety disorders and their offspring have reduced heart rate variability: implications for the health of children. PLoS One, 8(12), e83186. | HPA-axis / heart-rate variability |
| 33 | Eykens, H., Widjaja, D., Vanderperren, K., Taelman, J., Braeken, M., Otte, R., Van den Bergh, B., Van Huffel, S. (2012). Phase-rectified signal averaging for the quantification of the influence of prenatal anxiety on heart rate variability of babies. In: BIOSIGNALS 2012 Proceedings (163-168). Vilamoura, Portugal. | Venue: conference proceedings — BIOSIGNALS/SciTePress, double-blind peer-reviewed at paper level, but not published in a peer-reviewed journal |
| 34 | Widjaja, D., Taelman, J., Vandeput, S., Braeken, M.A.K.A., Otte, R.A., Van den Bergh, B., Van Huffel, S. (2010). ECG-Derived Respiration: Comparison and New Measures for Respiratory Variability. Proc. 34th Annual Computers in Cardiology (CinC 2010), 37, 149-152. | Venue: conference proceedings — Computing in Cardiology; conference acceptance based on abstract review only, full paper not independently peer-reviewed; not a peer-reviewed journal |
| 35 | Widjaja, D., Vandeput, S., Taelman, J., Braeken, M.A.K.A., Otte, R.A., Van den Bergh, B., Van Huffel, S. (2010). Accurate R Peak Detection and Advanced Preprocessing of Normal ECG for Heart Rate Variability Analysis. Proc. 37th Annual Computers in Cardiology (CinC 2010), 533-536. | Venue: conference proceedings — Computing in Cardiology; conference acceptance based on abstract review only, full paper not independently peer-reviewed; not a peer-reviewed journal |
| 36 | Taelman, J., Vandeput, S., Widjaja, D., Braeken, M., Otte, R., Van den Bergh, B., Van Huffel, S. (2010). Stress during pregnancy: Is the autonomic nervous system influenced by anxiety? Proc. 37th Annual Computing in Cardiology, 37, 725-728. | Venue: conference proceedings — Computing in Cardiology; conference acceptance based on abstract review only, full paper not independently peer-reviewed; not a peer-reviewed journal |

**Table S2. Leuven observed–imputed comparison and clinical threshold counts, both cohorts (reference: Dutch female community norms)**

| Cohort | Period | Data | Valid n | Mean (SD) | P25 (decile) | n ≥40 | % ≥40 | n ≥42 | % ≥42 | n ≥43 | % ≥43 |
|---|---|---|---|---|---|---|---|---|---|---|---|
| Leuven | 12–22 wk | Observed | 79 | 39.35 (8.86) | 32.00 (D4) | 34 | 43.0 | 28 | 35.4 | 27 | 34.2 |
| Leuven | 12–22 wk | Imputed | 86 | 39.29 (8.54) | 33.00 (D4) | 36 | 41.9 | 29 | 33.7 | 28 | 32.6 |
| Leuven | 23–31 wk | Observed | 69 | 34.84 (9.10) | 28.00 (D3) | 15 | 21.7 | 13 | 18.8 | 12 | 17.4 |
| Leuven | 23–31 wk | Imputed | 86 | 34.92 (8.43) | 30.00 (D3) | 18 | 20.9 | 15 | 17.4 | 13 | 15.1 |
| Leuven | 32–40 wk | Observed | 73 | 36.30 (8.21) | 30.00 (D3) | 24 | 32.9 | 16 | 21.9 | 14 | 19.2 |
| Leuven | 32–40 wk | Imputed | 86 | 36.45 (7.62) | 32.00 (D4) | 26 | 30.2 | 16 | 18.6 | 14 | 16.3 |
| Tilburg | 8–14 wk | Observed | 176 | 33.47 (8.97) | 27.00 (D2) | 35 | 19.9 | 29 | 16.5 | 27 | 15.3 |
| Tilburg | 15–22 wk | Observed | 170 | 32.15 (7.94) | 27.00 (D2) | 22 | 12.9 | 20 | 11.8 | 19 | 11.2 |
| Tilburg | 31–37 wk | Observed | 153 | 34.23 (8.83) | 29.00 (D3) | 30 | 19.6 | 26 | 17.0 | 23 | 15.0 |

*Note. Deciles are based on the female community reference distribution of Van der Ploeg et al. (1980). Scores of 42 and 43 fall within the same decile (D7); a score of 40 falls one decile lower (D6). Observed and imputed Leuven values are closely comparable at all three periods. Maternal distributions are based on 86 mothers (twin pairs counted once). In the reappraised Leuven publications, the high-anxiety group was defined by a score of 43 or higher (the ≥ 43 columns), except in Van den Bergh et al. (2006), which used its sample-specific 75th percentile of 43.5 (scores of 44 or higher). In the Tilburg publications, State anxiety was used as a predictor only in the emotional face–voice study; the other publications used the SCL-90. All predictors were analysed continuously in the primary analyses; these threshold counts are descriptive only.*

**Table S3. Cross-cohort statistical comparisons of maternal State-anxiety (Leuven vs. Tilburg)**

| Leuven period | Tilburg period | Mean diff. | Welch t (df) | p | Cohen's d | 95% CI for d |
|---|---|---|---|---|---|---|
| 12–22 weeks | 8–14 weeks | 5.82 | t(176.4) = 5.09 | <.001 | 0.66 | [0.40, 0.92] |
| 12–22 weeks | 15–22 weeks | 7.14 | t(160.2) = 6.47 | <.001 | 0.88 | [0.61, 1.15] |
| 23–31 weeks | — | — | — | — | — | No corresponding Tilburg measurement window; comparison not performed |
| 32–40 weeks | 31–37 weeks | 2.22 | t(198.5) = 2.04 | .043 | 0.26 | [−0.00, 0.53] |

*Note. Independent-samples Welch's t-tests (unequal variances), computed from the group means, SDs, and n's reported in Table 1. Positive mean differences indicate higher Leuven values. Cohen's d computed with pooled SD; 95% CI via the Hedges–Olkin standard error. Welch's t =* $(M1-M2)/\sqrt{(SD1^2/n1+SD2^2/n2)}$*, with df estimated via the Welch–Satterthwaite equation; Cohen's d =* $(M1-M2)/SDpooled$*, where* $SDpooled = \sqrt{[((n1-1)SD1^2+(n2-1)SD2^2)/(n1+n2-2)]}$*. The two early Tilburg windows are tested separately against Leuven's 12–22-week window rather than pooled, because they largely involve the same mothers assessed twice; pooling would inflate power by treating repeated measures as independent. No Tilburg measurement corresponds to Leuven's 23–31-week window. These are descriptive comparisons based on summary statistics, not individual-level data.*

## Table S4. Evidence Audit: Key Findings, Key Nulls, Effect Sizes, Uncertainty, and Evidential-Weight Tier for the Eleven Reappraised Publications

| Publication | Finding | Source statistic | ES | 95% CI | Method / precision note | Tier |
|---|---|---|---|---|---|---|
| Van den Bergh et al. (2005) | Faster RT, Encoding task (main effect) | F(1,52)=6.01, p=.018 | d=0.72 | [0.13, 1.32] | d reconstructed from F and reported group sizes | 1 |
| | More errors, target-absent (main effect) | F(1,52)=5.91, p=.019 | d=0.72 | [0.12, 1.31] | idem | 1 |
| | Key null: Stop task / externally cued inhibition (SSRT primary) | SSRT F(1,52)=1.31, p=.257; MRT F=2.40, p=.127; SDRT F=3.66, p=.061; go accuracy F=1.09, p=.302 | d≈0.34 (SSRT) | [-0.24, 0.92] | Approx. d from SSRT F and n; CI via Hedges–Olkin SE; MDE d≈0.83 | 1 |
| Van den Bergh et al. (2006) | Boys: MRT increase over time-on-task | F(1,53)=4.02, p=.05 | d=0.74 | [-0.00, 1.47] | d reconstructed from F and reported boys' group sizes (11 HA/23 LMA) | 1 |
| | Boys: SD-RT increase over time-on-task | F(1,53)=6.88, p<.01 | d=0.96 | [0.21, 1.72] | idem | 1 |
| Mennes et al. (2006) | Lower % correct, Shifting Attentional Set (main effect) | F(1,41)=5.34, p=.026 | d=0.72 | [0.09, 1.34] | d reconstructed from F and reported group sizes | 1 |
| | 3-way interaction anxiety×compatibility×shift | F(1,41)=5.78, p=.021 | d=0.75 | [0.12, 1.37] | idem | 1 |
| | Key null: working memory (N-back) | F(1,41)<0.13, ns | d<0.11 | — | Only an upper bound on F was reported; no exact ES/CI; MDE d≈0.87 | 1 |
| | Key null: externally cued inhibition (Go/NoGo) | F(1,41)<0.35, ns | d<0.18 | — | Only an upper bound on F was reported; no exact ES/CI; MDE d≈0.87 | 1 |
| | Key null: visual orienting/reorienting (Cued Attention) | F(1,41)<1.5, ns | d<0.38 | — | Only an upper bound on F was reported; no exact ES/CI; MDE d≈0.87 | 1 |
| Mennes et al. (2009) | Less optimal gambling/inhibition allocation (behaviour) | F(1,18)=31.75, p<.001 | d=2.47 | [1.35, 3.58] | d reconstructed from F and reported group sizes (8 HA/15 LMA) | 1 |
| | Larger frontal P2a (electrode Fp2) | F(1,18)=4.88, p<.05 | d=0.97 | [0.07, 1.87] | idem | 1 |
| | Key null: Go/NoGo behaviour and ERP (exogenous control) | No anxiety effect on Go/NoGo performance; Go and Nogo ERP SPM contrasts yielded no significant results | — | — | Exact inferential statistic for the null contrast not reported; MDE d≈1.23 | 1 |
| Mennes et al. (2020) | Endogenous/exogenous modulation absent in HA — right IFJ | F(1,15)=11.27, p<.01 | d=1.63 | [0.53, 2.73] | d reconstructed from F and reported group sizes (8 HA/9 LMA) | 1 |
| | — right inferior MFG | F(1,15)=16.94, p<.001 | d=2.00 | [0.83, 3.17] | idem | 1 |
| | — left inferior precentral sulcus | F(1,15)=13.18, p<.01 | d=1.76 | [0.64, 2.89] | idem | 1 |
| | — right inferior precentral sulcus | F(1,15)=6.80, p<.02 | d=1.27 | [0.22, 2.31] | idem | 1 |

| Publication | Finding | Source statistic | ES | 95% CI | Method / precision note | Tier |
|---|---|---|---|---|---|---|
| | Key null: brain activity related to mere response inhibition | No anxiety-related effects in response-inhibition activity or regions commonly involved in successful inhibition; no single test statistic reported | — | — | Spatial/functional null, not reducible to one group contrast; MDE d≈1.36 | 1 |
| Van den Bergh et al. (2025) | Lower Vocabulary score (age 28) | $F(1,46)=10.87$, $p=.002$; partial $\eta^2=.191$, 95% CI [.03,.374] | d=1.10 | [0.41, 1.79] | d reconstructed from F/group sizes; original partial $\eta^2$ and CI reported directly | 3 |
| | Key null: Perceptual Organization and Processing Speed | Significant overall MANOVA; follow-up analyses reported that only Vocabulary differed significantly | — | — | Separate univariate statistics/standardized ES for these null outcomes not reported; MDE d≈0.93 | 3 |
| | Lower RD, left sup. post. corona radiata (cluster 1) | peak Z=4.03, $p_{cluster}=.027$, N=48 | — | — | Voxel/cluster result; no standardized cluster-level ES recoverable from published summary | 3 |
| | Lower MD, same cluster (cluster 2) | peak Z=3.80, $p_{cluster}=.037$, N=48 | — | — | idem | 3 |
| | Clusters 3–12: exploratory, $p_{cluster}>.05$, incl. FISO/FA/AD/MK | $p_{cluster}=.054$–.434 | — | — | Not cluster-significant; reported as context only | 3 |
| | SLF mean-kurtosis – Vocabulary correlation | $r(48)=-.349$, $p=.015$ | r=−0.35 | [−0.60, −0.02] | Direct from publication; did not survive multiplicity correction | 3 |
| Turk et al. (2023) | MPFC–left IFG connectivity (continuous prenatal anxiety) | bivariate $r=-.30$, $p=.035$, N=49 | r=−0.30 | [−0.54, −0.02] | Fisher z-transformation | 3 |
| | MPFC–left IFG connectivity (group contrast, uncorrected for postnatal anxiety) | $t(45)=3.54$, $p<.001$ | d=1.15 | [0.47, 1.82] | d reconstructed from t and reported group sizes | 3 |
| | MPFC–left IFG: variance explained by prenatal anxiety, over covariates | $\Delta R^2=.20$ ($R^2_{full}=.28$ vs. $R^2_{cov}=.08$) | sr≈0.45 | — | $\sqrt{\Delta R^2}$ gives magnitude of semi-partial correlation for the added predictor | 3 |
| | Left PFC–left supramarginal gyrus connectivity (NBS) | $p_{uncorr}=.0015$, $p_{FDR}<.05$ | — | — | Permutation-based subnetwork result; no per-edge standardized ES reported | 3 |
| | Key null: global graph/network metrics | No HA–LMA differences in density, strength, global clustering, normalized clustering, global efficiency, normalized efficiency, or small-worldness; all $p>.05$, including covariate-adjusted analyses | — | — | Global-network comparison null; exact statistics in supplementary table, not summarized as one ES; MDE d≈0.91 | 3 |
| von Leupoldt et al. (2017) | Greater dyspnoea-unpleasantness slope (HA vs. LMA, from M/SD) | HA .20(.03) vs LMA .15(.08), $p<.01$ | d=0.70 | [−0.03, 1.43] | d from group means/SD (pooled SD) | 3 |
| | Correlation of unpleasantness slope with maternal stress (continuous) | Spearman $\rho=.34$, $p<.05$, N=40 | ρ=0.34 | [0.03, 0.59] | Fisher z CI used as approximation for ρ | 3 |

| Publication | Finding | Source statistic | ES | 95% CI | Method / precision note | Tier |
|---|---|---|---|---|---|---|
| | Key null: dyspnoea intensity slopes | Single-breath: HA .17(.06) vs LMA .15(.07), p=.53; five-breath: HA .20(.03) vs LMA .16(.08), p=.30 | d≈0.30; 0.56 | [-0.42,1.01]; [-0.17,1.29] | d from reported means/SD; wide CIs; MDE d≈1.02 | 3 |
| van den Heuvel, Donkers, et al. (2015) | Larger N250 to standard tones — maternal anxiety | F(1,77)=8.18, p=.005, $\eta^2p$=.096 | r=0.31 | [0.10, 0.50] | r=$\sqrt{\eta^2p}$; Fisher z CI | 1 |
| | Smaller N250 — maternal mindfulness | F(1,77)=8.50, p=.005, $\eta^2p$=.099 | r=−0.32 | [−0.50, −0.10] | idem | 1 |
| | Larger P150 — maternal mindfulness | F(1,77)=10.48, p=.002, $\eta^2p$=.130 | r=0.35 | [0.14, 0.53] | idem | 1 |
| | Key null: anxiety associations with rare deviants (ISI-deviant, white noise, novel) | No significant association between prenatal anxiety and any measured ERP amplitude for the three rare deviant types | — | — | Exact F values for these null tests not reported; continuous-predictor MDE \|r\|≈.31 | 1 |
| Otte et al. (2015) | P350: Auditory Emotion × maternal-anxiety interaction | F(1,80)=4.25, p=.043, $\eta^2p$=.050 | r=0.22 | [0.01, 0.42] | r=$\sqrt{\eta^2p}$; Fisher z CI | 2 |
| | Constraint/null: no robust visual-prime-dependent anxiety effect | P650 Auditory Emotion × Visual Prime × GenAnx: F(1,80)=3.167, p=.079, $\eta^2p$=.038; disappeared after postnatal-anxiety adjustment | r≈0.19 | [−0.02, 0.40] | Trend-level higher-order interaction; CI from $\eta^2p$ conversion; not robust to postnatal-anxiety adjustment | 2 |
| van den Heuvel et al. (2018b) | Larger LPP, neutral pictures, middle window (central) | Pearson r=−.242, p=.025, N=86 | r=−0.24 | [−0.43, −0.03] | Fisher z-transformation | 2 |
| | Larger LPP, neutral pictures, late window (central) | Pearson r=−.272, p=.011, N=86 | r=−0.27 | [−0.46, −0.06] | idem | 2 |
| | Key comparison null: pleasant and unpleasant picture LPP | No significant prenatal-anxiety associations across regions/windows; largest \|r\|=.206 (pleasant, central early, p=.057); unpleasant \|r\|max=.117 | r range −.206 to .040 | All CIs include 0; e.g. r=−.206: [−.40,.01] | Direct correlations; continuous-predictor MDE \|r\|≈.30 | 2 |

*Note. ES = Cohen's d for dichotomous HA–LMA contrasts, r or ρ for continuous predictors, and sr for a semi-partial correlation. Where d was reconstructed from a reported F(1,df) or t statistic, the reported HA and LMA group sizes were used ($d = \sqrt{F} \times \sqrt{(1/nHA + 1/nLMA)}$, equivalently $t \times \sqrt{(1/nHA + 1/nLMA)}$); 95% CIs for d use the Hedges–Olkin standard-error approximation. Because several source analyses were ANCOVAs or repeated-measures models, these reconstructed d values are approximate standardized contrasts rather than fully adjusted model-based effect sizes. For r, 95% CIs use Fisher's z transformation. A dash indicates that no defensible standardized ES or CI could be recovered from the published summary. Key nulls are included only when they carry an inferential role in the present synthesis; they are not equivalence tests. MDE values refer to the approximate 80%-power benchmarks in Table 5 and do not account for covariates or repeated-measures structure.*